\documentclass[12pt]{article}

\usepackage{amsfonts}
\usepackage{amsmath}
\usepackage{amsthm}
\usepackage{cite}
\usepackage{epsfig}
\usepackage{latexsym}
\usepackage{paralist}
\usepackage{fancyhdr}
\usepackage{graphicx}
\numberwithin{equation}{section}
\usepackage[vcentermath]{youngtab}
\usepackage{young}
\usepackage{ytableau}
\usepackage{etex}
\usepackage{braket}
\usepackage{float}
\usepackage{extarrows}
\usepackage{autobreak}

\usepackage{pict2e}   
\usepackage{animate}  
\usepackage{multirow}
\usepackage{bigdelim}
\usepackage{fancybox}
\usepackage{cals}

\def\be{\begin{equation}}
\def\ee{\end{equation}}
\def\bea{\begin{eqnarray}}
\def\eea{\end{eqnarray}}

\renewcommand{\thefootnote}{\fnsymbol{footnote}}

\begin{document}

\hfuzz=100pt
\title{{\Large \bf{3d $Spin(N)$ Seiberg dualities} }}
\date{}
\author{ Keita Nii$^a$\footnote{keita.nii@yukawa.kyoto-u.ac.jp}
%, 
%Yuta Sekiguchi$^a$\footnote{yuta@itp.unibe.ch}
% and others$^{c,d}$
}
\date{\today}

\maketitle

\thispagestyle{fancy}
%\rhead{****-**-**}
\cfoot{}
\renewcommand{\headrulewidth}{0.0pt}

\vspace*{-1cm}
\begin{center}
%  \spa{0.5} \\
$^{a}${{\it Center for Gravitational Physics}}
\\{{\it Yukawa Institute for Theoretical Physics, Kyoto University
}}
\\ {{\it Kitashirakawa Oiwake-cho, Sakyo-Ku, Kyoto, Japan}}  
%\\{{\it  Sidlerstrasse 5, CH-3012 Bern, Switzerland}}
%\\ {{\it  }}
 % \spa{0.5} \\
%$^b${{\it Department of Physics}}
%\\ {{\it Nagoya University, Nagoya 464-8602, Japan}}
%\spa{0.5}  \\
%$^c${{\it  }}
%\\ {{\it }}
%\spa{0.5}  \\
%$^d${{\it }}

\end{center}

\begin{abstract}
We study low-energy aspects of 3d $\mathcal{N}=2$ $Spin(N)$ gauge theories with matters in vector and (conjugate) spinor representations. Extending the construction of the 4d $\mathcal{N}=1$ $Spin(N)$ Seiberg duality, we find 3d magnetic dual descriptions with tree-level superpotentials slightly different from the 4d ones. We test various consistency checks including RG flows to known 3d dualities and supersymmetry enhancement deformation which leads to a 3d $\mathcal{N}=4$ duality between $SU(2)$ with three hypermultiplets and $U(1)$ with four hypermultiplets. 
\end{abstract}

\renewcommand{\thefootnote}{\arabic{footnote}}
\setcounter{footnote}{0}

\newpage
\tableofcontents 
%\clearpage

\newpage

%%%%%%%%%%%%%%%%%%%%%%%%%%%%%%%%%%%%%%%%%%%%%%%%%%%%%%%%%%
%%%%%%%%%%%%%%%%%%%%%%%%%%%%%%%%%%%%%%%%%%%%%%%%%%%%%%%%%%
\section{Introduction}
%%%%%%%%%%%%%%%%%%%%%%%%%%%%%%%%%%%%%%%%%%%%%%%%%%%%%%%%%%
%%%%%%%%%%%%%%%%%%%%%%%%%%%%%%%%%%%%%%%%%%%%%%%%%%%%%%%%%%
%[Duality]
Duality is a powerful tool of studying low-energy structures of strongly-coupled and non-perturbative gauge theories. In supersymmetric theories with four supercharges, it is known as ``Seiberg duality'' and it was discovered by Seiberg in a 4d $\mathcal{N}=1$ $SU(N)$ gauge theory with fundamental flavors \cite{Seiberg:1994bz, Seiberg:1994pq}. After this first discovery of Seiberg duality, similar dualities have been constructed in various dimensions, in various gauge groups and for various representations of matters \cite{Kutasov:1995ve, Kutasov:1995np, Intriligator:1995id, Intriligator:1995ne, Intriligator:1995ax} (for 3d examples, see \cite{Giveon:2008zn, Niarchos:2008jb, Aharony:1997gp, Kapustin:2011vz, Kim:2013cma, Park:2013wta, Nii:2014jsa, Hwang:2018uyj}). 
%[Recently] 
Recently, the low-energy aspects of the Coulomb moduli space in 3d $\mathcal{N}=2$ supersymmetric gauge theories were deeply understood \cite{Aharony:1997bx, deBoer:1997kr, Intriligator:2013lca, Aharony:2013dha, Benini:2011mf, Aharony:2014uya, Aharony:2015pla, Csaki:2014cwa, Amariti:2015kha, Nii:2018bgf} and it became possible to construct various 3d Seiberg-like dualities.

%[In this paper] 
In this paper, we tackle wth a problem of constructing the Seiberg dualities for 3d $\mathcal{N}=2$ $Spin(N)$ gauge theories with vector and spinor matters. In \cite{Aharony:2013kma, Aharony:2011ci}, the $so(N)$ Seiberg duality with vector matters was proposed, where the gauge group could be $(S)O(N)$ or $(S)pin(N)$. In connection with this development, this paper will generalize it by adding spinorial matters. Therefore, we here consider the $Spin(N)$ gauge groups. In 4d, the duality for the $Spin(N)$ gauge theory with spinor and vector matters is known \cite{Pouliot:1995zc, Pouliot:1995sk, Pouliot:1996zh, Kawano:1996bd, Cho:1997kr, Berkooz:1997bb, Maru:1998hp}. We will claim that the similar dualities can  hold also in 3d with a small modification of the magnetic superpotential. The 3d superpotential will include a monopole superpotential for a dressed Coulomb branch. This modification must be necessary because of the following reasons: In 4d, a linear combination of the $U(1)$ global symmetries is anomalous and explicitly broken due to a chiral anomaly. The 4d dualities only respect the unbroken symmetries. For instance, the operator matching is well behaved under the unbroken symmetries. However, in a 3d limit, the anomalous $U(1)$ symmetry is restored since there is no chiral anomaly. Therefore, a naive 4d Seiberg duality does not straightforwardly hold in 3d. In this paper, we will discuss how one can remedy the failure of the 4d $Spin(N)$ Seiberg dualities in 3d. In $Spin(7)$ cases, we will give a calculation of the superconformal indices and find a nice agreement under the duality.

%[organization]
The rest of this paper is organized as follows. 
In Section 2, we consider the $Spin(7)$ gauge theory and its dual. We will explain how to modify the magnetic superpotential in order to have a correct 3d duality. As a byproduct of the $Spin(7)$ duality, we also discuss the $Spin(6)$ and $Spin(5)$ dualities. 
Section 3, 4 and 5 are devoted to the generalization of our argument to $Spin(8)$, $Spin(9)$ and $Spin(10)$ groups, respectively. These dualities are related to each other via some complex mass deformations. 
In Section 6, we summarize our findings and discuss possible future problems.

%%%%%%%%%%%%%%%%%%%%%%%%%%%%%%%%%%%%%%%%%%%%%%%%%%%%%%%%%%
%%%%%%%%%%%%%%%%%%%%%%%%%%%%%%%%%%%%%%%%%%%%%%%%%%%%%%%%%%
\section{3d $Spin(7)$ Seiberg duality}
%%%%%%%%%%%%%%%%%%%%%%%%%%%%%%%%%%%%%%%%%%%%%%%%%%%%%%%%%%
%%%%%%%%%%%%%%%%%%%%%%%%%%%%%%%%%%%%%%%%%%%%%%%%%%%%%%%%%%
We begin our analysis from the $Spin(7)$ examples with spinor matters whose dimension is $\mathbf{8}$. In four spacetime dimensions, the corresponding dualities were studied in \cite{Pouliot:1995zc, Cho:1997kr}. We will see that similar dualities can hold in 3d as well by slightly modifying the magnetic superpotential. The electric description is a 3d $\mathcal{N}=2$ $Spin(7)$ gauge theory with $F$ vector matters $Q$ and a spinor matter $S$, which is completely the same as the 4d electric theory. There is no tree-level superpotential. Since there is no chiral anomaly in 3d, the global symmetry becomes $SU(F) \times U(1) \times U(1) \times U(1)_R$. This enhanced symmetry forbids the conventional 4d Seiberg duality from being valid in 3d as advocated in Section 1. Table \ref{Spin7electric} summarizes the quantum numbers of the elementary fields. The analysis of the Higgs branch is similar to the 4d one \cite{Pouliot:1995zc, Cho:1997kr, Csaki:1996zb, Grinstein:1998bu}. We can define the following operators
\begin{gather*}
M_{QQ}:=Q^2,~~~M_{SS}:=S^2 \\
P_3:=SQ^3S,~~~P_4:=SQ^4S,~~~B:=Q^7,
\end{gather*}
where the gauge indices of spinor matters are contracted in a symmetric way. On the second line, the color (and also flavor) indices of $Q$'s are anti-symmetrized. See the quantum numbers of these operators in Table \ref{Spin7electric}. The meson operators on the first line will become elementary fields in a magnetic theory.

A new ingredient of 3d $\mathcal{N}=2$ supersymmetric gauge theories comes from the Coulomb branch which is a set of flat directions for the scalars in a vector superfield. At a generic point of the $Spin(7)$ Coulomb brach, the gauge group is spontaneously broken to $U(1)^3$. For each $U(1)$ factor, we can introduce a Coulomb branch coordinate (monopole operator) by dualizing the $U(1)$ vector superfield into a chiral superfield. However, almost all the (classical) flat directions are quantum-mechanically unstable and lifted because a monopole-instanton could generate a runaway potential for the Coulomb branch \cite{Affleck:1982as, Aharony:1997bx, deBoer:1997kr}. In $so(N)$ gauge theories, only a few directions are non-perturbatively allowed and become exactly flat \cite{Aharony:2011ci, Aharony:2013kma, Nii:2018tnd, Nii:2018wwj, Nii:2019wxi}. In the present case, there is a single flat direction: The electric Coulomb branch operator, which is denoted by $Y_{SO(5)}$, corresponds to the spontaneous gauge symmetry breaking \cite{Slansky:1981yr, Feger:2012bs, Georgi:1999wka}
\begin{align}
so(7) & \rightarrow so(5) \times u(1) \\
\mathbf{7}  & \rightarrow \mathbf{5}_{0}+\mathbf{1}_{\pm 2}  \\
\mathbf{8}  & \rightarrow  \mathbf{4}_{1}+\mathbf{4}_{-1},
\end{align}
where the Coulomb branch coordinate $Y_{SO(5)}$ is obtained by dualizing the unbroken $U(1)$ vector superfield into a chiral superfield. We can see the stability of this flat direction as follows: The vacuum of the remaining $SO(5)$ gauge dynamics can be stable and supersymmetric by the massless components of the vector matters since monopoles from $SO(5)$ have too many fermion zero-modes to create a runaway potential for the Coulomb branch. In order to describe this flat direction, we need to introduce a monopole operator $Y_{SO(5)}$. Since the spinor matters are charged under the unbroken $U(1)$ subgroup, all the spinor components are massive along this branch. For $Spin(7)$ gauge theories only with spinors, the $SO(5)$ dynamics obtains a runaway potential $W=\sum_{i=1,2}  \frac{1}{Y_i}$ from the fundamental monopoles of $U(1)^2 \subset SO(5)$ and hence $Y_{SO(5)}$ is lifted. In addition to this operator, we can also turn on a dressed Coulomb branch
\begin{align}
Y_{SO(5)}^{Q} := Y_{SO(5)} \left( \mathbf{5}_{0} \right)^{5} \sim  Y_{SO(5)} Q^5,
\end{align}
which is available only for theories with $F \ge 5$ vectors. This operator is also known as a baryon-monopole operator \cite{Aharony:2013kma}. The another Coulomb branch operator $Z_{SO(3)}$ \cite{Aharony:2013kma, Nii:2018tnd}, which corresponds to the breaking $so(7) \rightarrow so(3) \times su(2) \times u(1)$ and will be discussed in more detail in Section \ref{Spin72spinors}, cannot be a flat direction since the low-energy $SU(2)$ gauge theory includes a single doublet and its origin of the moduli space is eliminated \cite{Aharony:1997bx}. Therefore, for the $Spin(7)$ gauge theory with more than one spinor can have the flat direction $Z_{SO(3)}$.

For small flavors $F \le 4$, we find confinement descriptions whose low-energy spectrum is given by the Coulomb and Higgs branch operators. From the symmetry argument in Table \ref{Spin7electric}, the effective superpotential is determined as
\begin{gather*}
W^{eff}_{F=4} = Y_{SO(5)} (M_{QQ}^4 M_{SS}^2 +M_{QQ}P_3^2 +P_{4}^2 ),~~~
W^{eff}_{F=3} =   \lambda \left[ Y_{SO(5)} (M_{QQ}^3 M_{SS}^2 +P_3^2) -1   \right] \\
W^{eff}_{F=2} =\frac{1}{Y_{SO(5)} M_{QQ}^2 M_{SS}^2},~~~W^{eff}_{F=1} =\left( \frac{1}{Y_{SO(5)} M_{QQ} M_{SS}^2} \right)^{\frac{1}{2}},~~~W^{eff}_{F=0} =  \left( \frac{1}{Y_{SO(5)} M_{SS}^2} \right)^{\frac{1}{3}}.
\end{gather*}
For $F=4$, the theory exhibits s-confinement \cite{Nii:2018tnd} where there is no singularities on the whole (classical) moduli space of vacua. For $F=3$, the theory has one quantum constraint between the Higgs and Coulomb branch coordinates, where the origin of the moduli space is not a part of the moduli space and we can remove a single coordinate. For $F \le 2$, the theory obtains runaway potentials and hence there is no stable supersymmetric vacuum. For $F \ge 5$, we cannot write down effective superpotential without singularities at the origin of the moduli space. Hence, we expect that the non-abelian Coulomb phase at the origin of the moduli space is described by a magnetic dual description for $F \ge 5$. In this paper, we will always find a magnetic description for $Spin(N)$ gauge theories when the number of vectors is greater than the s-confinement value of $F$.

\begin{table}[H]\caption{3d $\mathcal{N}=2$ $Spin(7)$ gauge theory with $F$ vectors and a spinor} 
\begin{center}
\scalebox{1}{
  \begin{tabular}{|c||c|c|c|c|c| } \hline
  &$Spin(7)$&$SU(F)$&$U(1)$&$U(1)$& $U(1)_R$  \\ \hline
$Q$&$\mathbf{7}$&${\tiny \yng(1)}$&1&0&$0$ \\   
 $S$&$\mathbf{8}$&1&0&1&$0$ \\  \hline 
$M_{QQ}:=Q^2$ &1&${\tiny \yng(2)}$&2&0&0  \\
$M_{SS}:=S^2$ &1&1&0&2&0  \\
$P_3:=SQ^3S$ $(F \ge 3)$&1&\scriptsize 3-th anti-symm.&3&2&0  \\
$P_4:=SQ^4S$ $(F \ge 4)$ &1&\scriptsize 4-th anti-symm.&4&2&0 \\  
$B:=Q^7$ $(F \ge7)$&1&\scriptsize 7-th anti. symm.&7&0&0  \\ \hline
$Y_{SO(5)}$&1&1&$-2F$&$-4$&$2F-6$ \\ 
$Y_{SO(5)}^Q:=Y_{SO(5)} Q^5$ $(F \ge 5)$&1&\scriptsize 5-th anti. symm.&$5-2F$&$-4$&$2F-6$  \\ \hline
  \end{tabular}}
  \end{center}\label{Spin7electric}
\end{table}

%%%%%%%%%%%%%%%%%%%%%%%%%%%%%%%%%%%%%%%%%%%%%%%%%%%%%%%%%%
The magnetic description is given by a 3d $\mathcal{N}=2$ $SU(F-3)$ gauge theory with $F$ fundamental quarks $q$, a single fundamental quark $q'$, a symmetric-bar tensor $\bar{s}$ and two gauge singlets $M_{QQ}$ and $M_{SS}$. The fundamental quarks are distinguished into $q$ and $q'$ by a tree-level superpotential
\begin{align}
W_{mag}=M_{QQ}\bar{s}qq+\bar{s}q'q' +M_{SS}Y^{dressed},
\end{align}
where the last term is a mass term between the meson and dressed Coulomb branch operators.   
The precise definition of $Y^{dressed}$ will be given below. The superpotential completely fixes all the quantum numbers of the magnetic elementary fields, which is summarized in Table \ref{Spin7magnetic}. The superpotential is very similar to \cite{Pouliot:1995zc} but the last term is not identical to the 4d one ($W_{mag}^{4d}$ instead including $M_{SS} \det \bar{s}$). We will shortly see that this difference is quite important to find a correct operator matching under the 3d version of the duality.

The magnetic Higgs branch is described by the following operators and the operator mapping is straightforward:
\begin{align}
P_3 \sim q^{F-3},~~~~P_4 \sim q^{F-4}q',~~~Y_{SO(5)} \sim \det \bar{s}.
\end{align}
Notice that the last operator $\det \bar{s}$, which was removed in the 4d $Spin(7)$ duality due to an $F$-flatness condition of $M_{SS}$, is now transformed into one of the Coulomb branch operators. Next, we consider the matching of the magnetic Coulomb branch to the rest of electric operators. 
The bare Coulomb branch, which is denoted by $Y_{SU(F-5)}^{bare}$, corresponds to the gauge symmetry breaking
\begin{align}
SU(F-3) & \rightarrow SU(F-5) \times U(1)_1 \times U(1)_2 \\ 
{\tiny \yng(1)}  & \rightarrow  {\tiny \yng(1)}_{\, 0,-2} +\mathbf{1}_{1,F-5}+\mathbf{1}_{-1,F-5}\\
{\tiny \overline{ \yng(2)}} & \rightarrow {\tiny \overline{ \yng(2)}}_{\, 0,4} +{\tiny \overline{ \yng(1)}}_{\,-1,-F+7}  +{\tiny \overline{ \yng(1)}}_{\, 1,-F+7} +\mathbf{1}_{-2,-2F+10} +\mathbf{1}_{2,-2F+10}+\mathbf{1}_{0,-2F+10}   \\
\mathbf{adj.} & \rightarrow  \mathbf{adj.}_{0,0}+\mathbf{1}_{0,0}+\mathbf{1}_{0,0}+   \mathbf{1}_{2,0}+\mathbf{1}_{-2,0} \nonumber \\&\qquad + {\tiny \yng(1)} _{\,-1,-F+3}+{\tiny \yng(1)}_{\, 1,-F+3}+{\tiny \overline{ \yng(1)}}_{\, 1,F-3}+{\tiny \overline{ \yng(1)}}_{\, -1,F-3},
\end{align}
where the adjoint representation corresponds to the vector (gaugino $\tilde{W}_\alpha$) superfield. The chiral superfield $Y_{SU(F-5)}^{bare}$ is obtained by dualizing the $U(1)_1$ vector superfield.   
Since the matter content of the magnetic theory is ``chiral'' (unbalanced numbers of fundamental and anti-fundamental matters), a mixed Chern-Simons term between $U(1)_1$ and $U(1)_2$ is generated via the integration of the massive components along the Coulomb branch \cite{Intriligator:2013lca, Csaki:2014cwa, Amariti:2015kha, Aharony:2015pla, Nii:2018bgf, Nii:2019ebv}. The bare Coulomb branch $Y_{SU(F-5)}^{bare}$ thus obtains a non-zero $U(1)_2$ charge proportional to the mixed CS term $-k_{eff}^{U(1)_1,U(1)_2}=-4(F-5)$ \cite{Intriligator:2013lca}. In order to describe the magnetic Coulomb branch in a gauge-invariant way, we need to define the so-called dressed monopole operators
\begin{align}
Y^{dressed}:= Y_{SU(F-5)}^{bare} \left(  {\tiny \overline{ \yng(2)}}_{\, 0,4} \right)^{F-5} \sim  Y_{SU(F-5)}^{bare} \bar{s}^{F-5},
\end{align}
which is neutral under the $U(1)_2$ symmetry as it should be. The color indices of $\bar{s}^{F-5}$ are contracted by two epsilon tensors of the unbroken $SU(F-5)$.
This dressed operator couples to the meson singlet $M_{SS}$ via the tree-level superpotential and then the $F$-flatness condition of $M_{SS}$ gets rid of the dressed operator $Y^{dressed}$ from the chiral ring elements.

\begin{table}[H]\caption{The magnetic dual of Table \ref{Spin7electric}} 
\begin{center}
\scalebox{0.9}{
  \begin{tabular}{|c||c|c|c|c|c| } \hline
  &$SU(F-3)$&$SU(F)$&$U(1)$&$U(1)$& $U(1)_R$  \\ \hline
$q$&${\tiny \yng(1)}$&${\tiny \overline{\yng(1)}}$&$\frac{3}{F-3}$&$\frac{2}{F-3}$&$0$ \\  
$q'$&${\tiny \yng(1)}$&1&$\frac{F}{F-3}$&$\frac{2}{F-3}$&$0$ \\  
 $\bar{s}$&${\tiny \overline{\yng(2)}}$&1&$\frac{-2F}{F-3}$&$\frac{-4}{F-3}$&$2$ \\  
$M_{QQ}$ &1&${\tiny \yng(2)}$&2&0&0  \\
$M_{SS}$ &1&1&0&2&0  \\  \hline
%$B:=Q^7$&1&7-th anti-symm. &7&0&0  \\
$Y_{SO(5)} \sim \det \bar{s}$&1&1&$-2F$&$-4$&$2F-6$  \\
$P_3 \sim q^{F-3}$ &1&\scriptsize 3-th anti-symm.&3&2&0  \\
$P_4 \sim q^{F-4}q'$ &1&\scriptsize 4-th anti-symm.&4&2&0 \\   \hline
\footnotesize $Y_{SU(F-5)}^{bare}$&\footnotesize $U(1)_2$: $-4(F-5)$&1&\footnotesize  $2F-\frac{4F}{F-3}$&\footnotesize  $4-\frac{2F+2}{F-3}$&\footnotesize  $-2F+12$ \\ 
\footnotesize  $Y^{dressed}:= Y_{SU(F-5)}^{bare} \bar{s}^{F-5}$&1&1&0&$-2$&2  \\  
\footnotesize  $B \sim Y^{bare}_{SU(F-5)} (\bar{s}q)^{F-7} \tilde{W}_\alpha^2$&1&\scriptsize 7-th anti. symm.&7&0&0  \\
\footnotesize  $Y_{SO(5)}^Q \sim Y^{bare}_{SU(F-5)} (\bar{s}^{F-5}\bar{s})(\bar{s}q)^{F-5}$&1&\scriptsize 5-th anti. symm.&$5-2F$&$-4$&$2F-6$  \\ \hline
  \end{tabular}}
  \end{center}\label{Spin7magnetic}
\end{table}

The non-trivial matching of the gauge-invariant operators comes from the baryons and the other dressed Coulomb branch operators. On the electric side, the baryon operator $B:=Q^7$ is available for $F \ge 7$ whereas there is naively no such operator on the magnetic side. In addition, the dressed Coulomb branch $ Y_{SO(5)}^Q:=Y_{SO(5)}Q^5$ is available for $F \ge 5$ on the electric side. In the magnetic description, these operators are defined by more exotic dressed Coulomb branch (or known as baryon-monopoles \cite{Aharony:2015pla}) as follows:
\begin{align}
B & \sim Y^{bare}_{SU(F-5)} {\tiny \overline{\yng(1)}}_{0,2}^{F-7}{\tiny \overline{\yng(1)}}_{1,F-3} {\tiny \overline{\yng(1)}}_{-1,F-3}  \sim Y^{bare}_{SU(F-5)} (\bar{s}q)^{F-7} \tilde{W}_\alpha \tilde{W}_\alpha \\
Y_{SO(5)}^Q &\sim  Y^{bare}_{SU(F-5)}  \left(  {\tiny \overline{ \yng(2)}}_{\, 0,4}^{F-5} \mathbf{1}_{0,-2F+10} \right)  {\tiny \overline{\yng(1)}}_{0,2}^{F-5}  \sim Y^{bare}_{SU(F-5)} (\bar{s}^{F-5}\bar{s})(\bar{s}q)^{F-5},
\end{align}
where two $\tilde{W}_\alpha$'s represent gaugino contributions ${\tiny \overline{\yng(1)}}_{1,F-3} $ and ${\tiny \overline{\yng(1)}}_{-1,F-3}$. See \cite{Aharony:2015pla, Csaki:2014cwa, Amariti:2015kha, Kim:2013cma, Nii:2018bgf, Nii:2019ebv, Nii:2019qdx} for studies on the dressed Coulomb branch in other gauge theories. We listed the quantum numbers of these operators in Table \ref{Spin7magnetic} and this confirms the operator identification above.

%%%%%%%%%%%%%%%%%%%%%%%%%%%%%%%%%%%%%%%%%%%%%%%%%%%%%%%%%%
As a consistency check of our duality proposal, we first test the superconformal indices \cite{Bhattacharya:2008bja, Kim:2009wb, Imamura:2011su, Kapustin:2011jm} for the $F=5$ case. The dual gauge group becomes $SU(2)$. By using the localization technique developed in \cite{Pestun:2007rz, Kapustin:2009kz, Hama:2010av}, the superconformal indices of the electric and magnetic descriptions for $F=5$ are computed as

\footnotesize \begin{align}
 I_{F=5}&=1+\sqrt{x} \left(\frac{1}{t^{10} u^4}+15 t^2+u^2\right)+x \left(\frac{1}{t^{20} u^8}+\frac{1}{t^{10} u^2}+\frac{15}{t^8 u^4}+120 t^4+15 t^2 u^2+u^4\right)+10 t^3 u^2 x^{5/4}\nonumber \\ \nonumber 
&+x^{3/2} \left(\frac{1}{t^{30} u^{12}}+\frac{1}{t^{20} u^6}+\frac{15}{t^{18} u^8}+\frac{1}{t^{10}}+\frac{15}{t^8 u^2}+\frac{120}{t^6 u^4}+680 t^6+125 t^4 u^2+15 t^2 u^4+u^6\right)  \\ \nonumber
&+x^{7/4} \left(\frac{10}{t^7 u^2}+\frac{1}{t^5 u^4}+150 t^5 u^2+10 t^3 u^4\right)  +x^2 \left(\frac{1}{t^{40} u^{16}}+\frac{1}{t^{30} u^{10}}+\frac{15}{t^{28} u^{12}}+\frac{1}{t^{20} u^4}+\frac{15}{t^{18} u^6}+\frac{120}{t^{16} u^8} \right.  \\ 
&\qquad \left.+\frac{u^2}{t^{10}}+3060 t^8+\frac{15}{t^8}+755 t^6 u^2+\frac{120}{t^6 u^2}+125 t^4 u^4+\frac{680}{t^4 u^4}+15 t^2 u^6+u^8-26\right)  +\cdots,
\end{align}
\normalsize

\noindent where we set the r-charges to be $r_Q=r_S=\frac{1}{4}$ for simplicity but one can use any r-charges to observe this agreement. The parameters $t$ and $u$ are the fugacities for the two $U(1)$ symmetries. We observed this agreement of the indices up to $O(x^2)$. The interpretation of each term is in order: The second term $\sqrt{x} \left(\frac{1}{t^{10} u^4}+15 t^2+u^2\right)$ corresponds to a sum of three operators $Y_{SO(5)} + M_{QQ} +M_{SS}$. The coefficient represents a number of components. The fourth term $10 t^3 u^2 x^{5/4}$ is interpreted as $P_3$. $P_4$ corresponds to $5t^4 u^2 x^{3/2}$. The dressed Coulomb branch $Y_{SO(5)}^Q$ appears as $\frac{x^{7/4}}{t^5 u^4}$. The other terms are the symmetric product of these operators or fermion contributions.

As a second consistency check of the duality between Table \ref{Spin7electric} and Table \ref{Spin7magnetic}, let us consider the case with $F=4$ where the magnetic gauge group is vanishing and then we can expect that the electric theory exhibits an s-confinement phase. As studied in \cite{Nii:2018tnd}, the $Spin(7)$ gauge theory with four vectors and a single spinor shows s-confinement. 

We can also derive the 3d $G_2$ duality known in \cite{Nii:2019wjz}. On the electric side, we introduce a non-zero vev to $\braket{M_{SS}}=v^2$ which breaks the gauge group into $G_2$. The low-energy description becomes a 3d $\mathcal{N}=2$ $G_2$ gauge theory with $F$ fundamental matters. On the magnetic side, the gauge group is unchanged but the superpotential becomes
\begin{align}
W_{mag} =M_{QQ} \bar{s}qq +\bar{s} q'q' +v^2Y^{dressed},
\end{align}
which is completely the same as the dual description proposed in \cite{Nii:2019wjz} and the vev $v^2$ can be absorbed into the definition of $Y^{dressed}$.

By introducing a complex mass to the spinor field, we can obtain the $Spin(7)$ Seiberg duality with vectors and no spinor \cite{Aharony:2013kma}. On the electric side, the mass term $W=m M_{SS}$ just integrates out the spinor matter. On the magnetic side, the complex mass leads to the higgsing $Y^{dressed} \neq 0$. Since $Y^{dressed}$ is a composite operator consisting of the Higgs and Coulomb branch fields, the gauge group is spontaneously broken in a twofold way
\begin{align}
SU(F-3) &\rightarrow S(U(F-5) \times U(1)_1 \times U(1)_2)  \nonumber \\
&\rightarrow S(O(F-5) \times U(1)_1 ) \sim S(O(F-5) \times O(2)),
\end{align}
where the first breaking is induced by $Y^{bare}_{SU(F-5)}$ and the second breaking is caused by $\bar{s}^{F-5}$. By doing the gauge transformation of $SU(F-5)$, we can choose the vev of $\bar{s}$ as an identity matrix. Due to the vev of $\bar{s}$, $q'$ becomes massive.  
The Affleck-Harvey-Witten type superpotential \cite{Affleck:1982as} is generated by the $U(1)_1$ factor and the resulting superpotential becomes $W=M_{QQ}qq +Y \tilde{Y}$, where $\tilde{Y}$ is a Coulomb branch of $O(F-5)$. Since the $O(2)$ gauge dynamics is translated into a chiral superfield $Y$, the low-energy gauge group becomes $O(F-5)$. The singlet $Y$ and the Coulomb branch operator $\tilde{Y}$ are charged under the $\mathbb{Z}_2 \subset O(F-5)$. This is precisely the duality proposed in \cite{Aharony:2013kma}.

We can also turn on a complex mass to the vector matters and test the flow of the duality. As an example, let us introduce a mass deformation to a single flavor, say $\Delta W= m Q_{F}Q_F$. On the electric side, the vector matter $Q_F$ is just integrated out. On the magnetic side, the mass term leads to the higgsing $\braket{\bar{s}q^Fq^F}=-m$ which breaks the gauge group into $SU(F-4)$. As a result, the duality is preserved with a reduction of $F$.

%%%%%%%%%%%%%%%%%%%%%%%%%%%%%%%%%%%%%%%%%%%%%%%%%%%%%%%%%%
By introducing a non-zero vev with $\mathrm{rank} \braket{M_{QQ}} =r$, we can derive the $Spin(7-r)$ duality with some spinor matters. Here, we only consider the $r=1$ and $r=2$ cases. On the electric side, $\mathrm{rank} \braket{M_{QQ}} =1$ breaks the gauge group to $Spin(6)$ and the low-energy description becomes a 3d $\mathcal{N}=2$ $Spin(6)$ gauge theory with $F$ vectors, one spinor and one conjugate spinor, where we shifted $F$ to $F+1$. 
On the magnetic side,  $\mathrm{rank} \braket{M_{QQ}} =1$ decomposes $F+1$ quarks into $F$ quarks and a single quark which is combined with $q'$ into $q_+$ and $q_-$. As a result, the magnetic description becomes a 3d $\mathcal{N}=2$ $SU(F-2)$ gauge theory with $F$ quarks $q$, two quarks $q_{\pm}$, one symmetric-bar tensor $\bar{s}$ and two gauge-singlets $M_{QQ}$ and $M_{SS}$. The theory has a tree-level superpotential  
\begin{align}
W_{mag}=M_{QQ}\bar{s}qq +\bar{s}q_+ q_- + M_{S\bar{S}} \tilde{Y}^{dressed},
\end{align}
which completely fixes the charge assignment of the dual elementary field as listed in Table \ref{Spin6}. The mapping of the magnetic Higgs branch operators is easily obtained 
\begin{gather*}
P_2 \sim q^{F-2},~~~P_3 \sim q^{F-3} q_+,~~~\bar{P}_3 \sim q^{F-3}q_-,~~~P_4 \sim q^{F-4} q_+q_-,~~~Y_{SO(4)} \sim \bar{s}^{F-2}.
\end{gather*}
When the bare Coulomb branch $\tilde{Y}^{bare}_{SU(F-4)}$ in the magnetic theory obtains a non-zero vev, the gauge group is spontaneously broken to $SU(F-2) \rightarrow SU(F-4) \times U(1)_1 \times U(1)_2$. The gauge-invariant monopoles are similarly defined by
\begin{align}
\tilde{Y}^{dressed} &:=\tilde{Y}^{bare}_{SU(F-4)} \bar{s}^{F-4} \\
B&\sim \tilde{Y}^{bare}_{SU(F-4)} (\bar{s}q)^{F-6} \tilde{W}_\alpha^2 \\
Y_{SO(4)}^{Q} &\sim  \tilde{Y}^{bare}_{SU(F-4)} \bar{s}^{F-4}\bar{s} (\bar{s}q)^{F-4}, 
\end{align}
whose global charges are consistent with the duality in Table \ref{Spin6}.

As a consistency check of the duality shown in Table \ref{Spin6}, we study the superconformal indices of the electric and magnetic theories. We computed the indices for the $F=4$ case and observed a nice agreement up to $O(x^2)$. The result is given by

\footnotesize
\begin{align}
I_{F=4}&=1+x^{1/2} \left(t u+10 v^2\right)+x \left(\frac{1}{t^2 u^2 v^8}+t^2 u^2+16 t u v^2+55 v^4\right) \nonumber \\ & +x^{5/4} \left(4 t^2 v^3+4 u^2 v^3\right)  +x^{3/2} \left(t^3 u^3+\frac{10}{t^2 u^2 v^6}+16 t^2 u^2 v^2+\frac{1}{t u v^8}+116 t u v^4+220 v^6\right) \nonumber \\
&+x^{7/4} \left(4 t^3 u v^3+40 t^2 v^5+4 t u^3 v^3+40 u^2 v^5\right) \nonumber \\
&+x^2 \left(\frac{1}{t^4 u^4 v^{16}}+t^4 u^4+16 t^3 u^3 v^2+136 t^2 u^2 v^4+\frac{56}{t^2 u^2 v^4}+560 t u v^6+\frac{16}{t u v^6}+715 v^8+\frac{1}{v^8}-18\right)+\cdots,
\end{align}
\normalsize

\noindent where $v$ is a fugacity for the $U(1)$ symmetry rotating vectors $Q$ while $t$ and $u$ count the numbers of the spinor and conjugate spinor. We set the r-charges as $r_{Q}=r_S=r_{\bar{S}}=\frac{1}{4}$ for simplicity. The second term $x^{1/2} \left(t u+10 v^2\right)$ corresponds to the two mesons $M_{S\bar{S}} + M_{QQ}$. The fourth term $x^{5/4} \left(4 t^2 v^3+4 u^2 v^3\right) $ is a contribution of $P_3$ and $\bar{P}_3$. The composites $P_2$ and $P_4$ are identified with $6tuv^2x$ and $tuv^4x^{3/2}$, respectively. 
The bare Coulomb branch $Y_{SO(4)}$ is represented as $\frac{x}{t^2 u^2 v^8}$ while the dressed one $Y_{SO(4)}^{Q}$ corresponds to $\frac{x^2}{t^2 u^2 v^4}$. Note that the lowest contribution from the state with a GNO charge $(1,1,0)$ could be classically regarded as a Coulomb branch $Z_{SU(2)}$ whose breaking pattern corresponds to $Spin(6) \rightarrow SU(2) \times U(1)\times U(1)$. This seems to contradict our analysis of the Coulomb branch. However, this should be regarded as the product $Y_{SO(4)} M_{S \bar{S}}$ and is consistent with our analysis since the low-energy $SU(2)$ dynamics along $Z_{SU(2)}$ is unstable.

\begin{table}[H]\caption{3d $\mathcal{N}=2$ $Spin(6)$ gauge theory with $F$ vectors, a spinor and a conjugate spinor and its $SU(F-2)$ dual description. The top table corresponds to the electric theory while the bottom one is magnetic. } 
\begin{center}
\scalebox{0.95}{
  \begin{tabular}{|c||c|c|c|c|c|c| } \hline
  &$Spin(6)$&$SU(F)$&$U(1)$&$U(1)$&$U(1)$& $U(1)_R$  \\ \hline
$Q$&$\mathbf{6}$&${\tiny \yng(1)}$&1&0&0&$0$ \\   
 $S$&$\mathbf{4}$&1&0&1&0&$0$ \\  
 $\bar{S}$&$\bar{\mathbf{4}}$&1&0&0&1&$0$ \\  \hline
$M_{QQ}:=QQ$ &1&${\tiny \yng(2)}$&2&0&0&0  \\
 $M_{S\bar{S}}:=S\bar{S}$&1&1&0&1&1&0  \\
 $P_2:=SQ^2\bar{S}$&1&${\tiny \yng(1,1)}$&2&1&1&0  \\
 $P_3:=SQ^3S$&1&\scriptsize 3-th anti-symm.&3&2&0&0  \\
 $\bar{P}_3:=\bar{S}Q^3\bar{S}$&1&\scriptsize 3-th anti-symm.&3&0&2&0  \\
$P_4:=SQ^4\bar{S}$&1&\scriptsize 4-th anti-symm.&4&1&1&0  \\
$B:=Q^6$&1&\scriptsize 6-th anti-symm.&6&0&0&0  \\ \hline
$Y_{SO(4)}$&1&1&$-2F$&$-2$&$-2$&$2F-4$  \\
$Y^Q_{SO(4)}:=Y_{SO(4)} Q^4$&1&\scriptsize 4-th anti-symm.&$4-2F$&$-2$&$-2$&$2F-4$  \\ \hline \hline
&$SU(F-2)$&$SU(F)$&$U(1)$&$U(1)$&$U(1)$& $U(1)_R$  \\ \hline 
$q$&${\tiny \yng(1)}$&${\tiny \overline{\yng(1)}}$&$\frac{2}{F-2}$&$\frac{1}{F-2}$&$\frac{1}{F-2}$&0  \\
$q_+$&${\tiny \yng(1)}$&1&$\frac{F}{F-2}$& \scriptsize$1+\frac{1}{F-2}$&\scriptsize $-1+\frac{1}{F-2}$&0  \\
$q_-$&${\tiny \yng(1)}$&1&$\frac{F}{F-2}$&\scriptsize $-1+\frac{1}{F-2}$&\scriptsize $1+\frac{1}{F-2}$&0 \\
$\bar{s}$&${\tiny \overline{\yng(2)}}$&1&$-\frac{2F}{F-2}$&$-\frac{2}{F-2}$&$-\frac{2}{F-2}$&2 \\
$M_{QQ}$ &1&${\tiny \yng(2)}$&2&0&0&0  \\
 $M_{S\bar{S}}$&1&1&0&1&1&0  \\  \hline
 $P_2 \sim q^{F-2}$&1&${\tiny \yng(1,1)}$&2&1&1&0  \\
 $P_3 \sim q^{F-3} q_+$&1&\scriptsize 3-th anti-symm.&3&2&0&0  \\
 $\bar{P}_3 \sim q^{F-3}q_-$&1&\scriptsize 3-th anti-symm.&3&0&2&0  \\
$P_4 \sim q^{F-4} q_+q_-$&1&\scriptsize 4-th anti-symm.&4&1&1&0  \\
$Y_{SO(4)} \sim \bar{s}^{F-2}$&1&1&$-2F$&$-2$&$-2$&$2F-4$  \\ \hline
\scriptsize $\tilde{Y}^{bare}_{SU(F-4)} $&\tiny $U(1)_2$: $-4(F-4)$&1&\scriptsize $2F-\frac{4F}{F-2}$&\scriptsize $2-\frac{F+2}{F-2}$&\scriptsize $2-\frac{F+2}{F-2}$&\scriptsize $10-2F$  \\
\scriptsize $\tilde{Y}^{dressed}:=\tilde{Y}^{bare}_{SU(F-4)} \bar{s}^{F-4}$ &1&1&0&$-1$&$-1$&2  \\
\scriptsize $B\sim \tilde{Y}^{bare}_{SU(F-4)} (\bar{s}q)^{F-6} \tilde{W}_\alpha^2$&1&\scriptsize 6-th anti-symm.&6&0&0&0 \\  
\scriptsize $ Y_{SO(4)}^Q \sim  \tilde{Y}^{bare}_{SU(F-4)} \bar{s}^{F-4}\bar{s} (\bar{s}q)^{F-4} $&1&\scriptsize 4-th anti-symm.&$4-2F$&$-2$&$-2$&$2F-4$  \\ \hline 
  \end{tabular}}
  \end{center}\label{Spin6}
\end{table}

%%%%%%%%%%%%%%%%%%%%%%%%%%%%%%%%%%%%%%%%%%%%%%%%%%%%%%%%%%
Finally, we discuss the $Spin(5)$ flat direction by introducing a non-zero expectation value such that $\mathrm{rank}\, \braket{M_{QQ}}=2$, which breaks $Spin(7)$ to $Spin(5)$. The electric description becomes a 3d $\mathcal{N}=2$ $Spin(5)$ gauge theory with $F$ vectors $Q$ and two spinors $S$, which is equivalent to a $USp(4)$ with $F$ anti-symmetric tensors and two fundamental matters. The non-abelian global symmetry is then enhanced to $SU(F) \times SU(2)$. The electric Higgs branch is described by
\begin{gather*}
M_{QQ}:=QQ,~~~M_{SS}:=SS,~~~P_{1A}:=SQS,~~~P_{2S}:=SQ^2S \\
P_{3S}:=SQ^3S,~~~P_{4A}:=SQ^4S,~~~B:=Q^5,
\end{gather*}
whose quantum numbers are summarized in Table \ref{Spin5}. In $P_{1A}$ and $P_{4A}$, two spinors are anti-symmertrized.  
The electric Coulomb branch, denoted by $Y_{SO(3)}$, corresponds to the gauge symmetry breaking $so(5) \rightarrow so(3) \times u(1)$. We can also define the baryon-monopole $Y^Q_{SO(3)}:=Y_{SO(3)} Q^3$.

The magnetic side becomes a 3d $\mathcal{N}=2$ $SU(F-1)$ gauge theory with $F+1$ fundamental matters, a symmetric-bar tensor and two gauge-singlet mesons $M_{QQ}$ and $M_{SS}$. The magnetic theory has a tree-level superpotential
\begin{align}
W_{mag} =M_{QQ} \bar{s}qq + \bar{s}q'q' +M_{SS}\tilde{Y}^{dressed},
\end{align}
which decomposes fundamental matters into $q$ and $q'$, reducing the global symmetry into $SU(F) \times SO(3) \times U(1) \times U(1)$. The charge assignment of the magnetic fields is completely fixed by this superpotential as listed in Table \ref{Spin5}. The Higgs branch operators on the magnetic side, which are not truncated by the $F$-flatness conditions, are mapped under the duality as follows:
\begin{gather*}
P_{1A}\sim q^{F-1},~~~P_{2S} \sim q^{F-2}q',~~~P_{3S} \sim q^{F-3} q'^2  \\
P_{4S} \sim q^{F-4} q'^3,~~~Y_{SO(3)} \sim \bar{s}^{F-1}.
\end{gather*}
Notice that the operator $\bar{s}^{F-1}$ is mapped to the electric Coulomb branch $Y_{SO(3)}$. We can also find the correct matching of the magnetic Coulomb branch: The bare Coulomb branch $\tilde{Y}^{bare}_{SU(F-3)}$ corresponds to the breaking $SU(F-1) \rightarrow SU(F-3) \times U(1)_1 \times U(1)_2$ and its $U(1)_2$ charge is computed as $-4(F-3)$. In order to construct gauge-invariants, we need to define baryon-monopole operators
\begin{align}
\tilde{Y}^{dressed} &:=\tilde{Y}^{bare}_{SU(F-3)} \bar{s}^{F-3}  \\
B &:=  \tilde{Y}^{bare}_{SU(F-3)} (\bar{s}q)^{F-5} \tilde{W}_\alpha^2,
\end{align}
where the first operator $\tilde{Y}^{dressed} $ is eliminated due to the superpotential and the second one is identified with the baryon $B$. The quantum numbers of these operators are summarized in Table \ref{Spin5}. 

\begin{table}[H]\caption{3d $\mathcal{N}=2$ $Spin(5)$ gauge theory with $F$ vectors and two spinors and its dual description. The top table shows an electric description while the bottom one corresponds to the magnetic description.} 
\begin{center}
\scalebox{0.99}{
  \begin{tabular}{|c||c|c|c|c|c|c| } \hline
  &$Spin(5)$&$SU(F)$&$SU(2)$&$U(1)$&$U(1)$& $U(1)_R$  \\ \hline
$Q$&$\mathbf{5}$&${\tiny \yng(1)}$&1&1&0&$0$ \\   
 $S$&$\mathbf{4}$&1&${\tiny \yng(1)}$&0&1&$0$ \\  \hline
$M_{QQ}:=QQ$ &1&${\tiny \yng(2)}$&1&2&0&0  \\
 $M_{SS}:=SS$&1&1&1&0&2&0  \\
 $P_{1A}:=SQS$&1&${\tiny \yng(1)}$&1&1&2&0 \\
 $P_{2S}:=SQ^2S$&1&${\tiny \yng(1,1)}$&${\tiny \yng(2)}$&2&2&0 \\
 $P_{3S}:=SQ^3S$&1&\scriptsize 3-th anti-symm.&${\tiny \yng(2)}$&3&2&0 \\
 $P_{4A}:=SQ^4S$&1&\scriptsize 4-th anti-symm.&1&4&2&0 \\
$B:=Q^5$&1&\scriptsize 5-th anti-symm.&1&5&0&0  \\ \hline
$Y_{SO(3)}$&1&1&1&$-2F$&$-4$&$2F-2$  \\
$Y^Q_{SO(3)}:=Y_{SO(3)} Q^3$&1&\scriptsize 3-th anti-symm.&1& $3-2F$&$-4$&$2F-2$  \\ \hline \hline
&$SU(F-1)$&$SU(F)$&$SO(3)$&$U(1)$&$U(1)$& $U(1)_R$  \\ \hline 
$q$&${\tiny \yng(1)}$&${\tiny \overline{\yng(1)}}$&1&$\frac{1}{F-1}$&$\frac{2}{F-1}$&0  \\
$q'$&${\tiny \yng(1)}$&1&${\tiny \yng(1)}$&$\frac{F}{F-1}$&$\frac{2}{F-1}$&0  \\
$\bar{s}$&${\tiny \overline{\yng(2)}}$&1&1&$-\frac{2F}{F-1}$&$-\frac{4}{F-1}$& 2 \\ 
  $M_{QQ}$ &1&${\tiny \yng(2)}$&1&2&0&0  \\
 $M_{SS}$&1&1&0&0&2&0  \\ \hline
  $P_{1A}\sim q^{F-1}$&1&${\tiny \yng(1)}$&1&1&2&0 \\
 $P_{2S} \sim q^{F-2}q' $&1&${\tiny \yng(1,1)}$&${\tiny \yng(1)}$&2&2&0 \\
 $P_{3S} \sim q^{F-3} q'^2 $&1&\scriptsize 3-th anti-symm.&${\tiny \yng(1)}$&3&2&0 \\
 $P_{4S} \sim q^{F-4} q'^3$&1&\scriptsize 4-th anti-symm.&1&4&2&0 \\
 $Y_{SO(3)} \sim \bar{s}^{F-1}$&1&1&1&$-2F$&$-4$&$2F-2$  \\ \hline
\footnotesize $\tilde{Y}^{bare}_{SU(F-3)} $ &\scriptsize $U(1)_2$: $-4(F-3)$&1&1&\scriptsize $2F-\frac{4F}{F-1}$&\scriptsize $2-\frac{8}{F-1}$&\scriptsize $8-2F$  \\ 
\footnotesize $\tilde{Y}^{dressed}:=\tilde{Y}^{bare}_{SU(F-3)} \bar{s}^{F-3}$ &1&1&1&$0$&$-2$&2  \\
\footnotesize $B\sim \tilde{Y}^{bare}_{SU(F-3)} (\bar{s}q)^{F-5} \tilde{W}_\alpha^2$&1&\scriptsize 5-th anti-symm.&1&5&0&0 \\  \hline
  \end{tabular}}
  \end{center}\label{Spin5}
\end{table}

For $F=2$, the electric $Spin(5) =USp(4)$ gauge theory exhibits s-confinement \cite{Benvenuti:2018bav, Amariti:2018wht, Nii:2019ebv}. This is consistent with our duality since the dual gauge group vanishes when $F=2$. This serves as a quick consistency check of the duality. As another test of the duality, we can check the superconformal indices for the $F=3$ case. By using the localization technique \cite{Pestun:2007rz, Kapustin:2009kz, Hama:2010av}, we find that both the electric and magnetic descriptions give a consistent result  

\small 
\begin{align}
I_{F=3} &= 1+x^{2/3} \left(\frac{1}{t^6 u^4}+6 t^2+u^2\right)+3 t u^2 x+x^{4/3} \left(\frac{1}{t^{12} u^8}+\frac{1}{t^6 u^2}+\frac{6}{t^4 u^4}+21 t^4+15 t^2 u^2+u^4\right) \nonumber \\
&+x^{5/3} \left(\frac{3}{t^5 u^2}+\frac{1}{t^3 u^4}+21 t^3 u^2+3 t u^4\right)  \nonumber \\
&+x^2 \left(\frac{1}{t^{18} u^{12}}+\frac{1}{t^{12} u^6}+\frac{6}{t^{10} u^8}+56 t^6+\frac{1}{t^6}+75 t^4 u^2+\frac{6}{t^4 u^2}+21 t^2 u^4+\frac{21}{t^2 u^4}+u^6-13\right)  \nonumber \\
&+x^{7/3} \left(\frac{3}{t^{11} u^6}+\frac{1}{t^9 u^8}+81 t^5 u^2+\frac{3}{t^5}+45 t^3 u^4+\frac{10}{t^3 u^2}+3 t u^6+\frac{6}{t u^4}-\frac{3 u^2}{t}-12 t\right)  \nonumber \\
&+x^{8/3} \left(\frac{1}{t^{24} u^{16}}+\frac{1}{t^{18} u^{10}}+\frac{6}{t^{16} u^{12}}+\frac{1}{t^{12} u^4}+\frac{6}{t^{10} u^6}+\frac{21}{t^8 u^8}+126 t^8-\frac{9}{t^6 u^4}   \right. \nonumber  \\ & \left. +245 t^6 u^2+\frac{u^2}{t^6}+150 t^4 u^4+\frac{6}{t^4}+21 t^2 u^6+\frac{15}{t^2 u^2}-81 t^2+u^8+\frac{56}{u^4}-36 u^2\right)   \nonumber \\ 
&+x^3 \left(\frac{3}{t^{17} u^{10}}+\frac{1}{t^{15} u^{12}}+\frac{3}{t^{11} u^4}+\frac{10}{t^9 u^6}+\frac{6}{t^7 u^8}+231 t^7 u^2-\frac{3}{t^7 u^2}+240 t^5 u^4-\frac{9}{t^5 u^4}    \right. \nonumber  \\ &  \qquad  \left. +\frac{3 u^2}{t^5}+55 t^3 u^6-76 t^3+\frac{10}{t^3}+3 t u^8-\frac{3 u^4}{t}+\frac{21 t}{u^4}-81 t u^2+\frac{21}{t u^2}\right)+\cdots, 
\end{align}
\normalsize

\noindent where $t$ and $u$ are the fugacities for the two $U(1)$ global symmetries. The r-charges are set to be $r_{Q}=r_{S}=\frac{1}{4}$ for simplicity. The operator interpretation of each term is easily obtained: In the second term $x^{2/3} \left(\frac{1}{t^6 u^4}+6 t^2+u^2\right)$, there are three contributions $Y_{SO(3)}+M_{QQ}+M_{SS}$, where the Coulomb branch $Y_{SO(3)}$ comes from the state with a non-zero GNO charge. The third term $3 t u^2 x$ corresponds to $P_{1A}$ and its coefficient correctly explains the representation under the non-abelian global symmetry. The composites $P_{2S}$ and $P_{3S}$ are represented as $9t^2 u^2x^{4/3}$ and $3 t^3 u^2 x^{5/3}$, respectively. The dressed Coulomb branch $Y_{SO(3)}^Q$ appears as $\frac{x^{5/3}}{t^3 u^4}$. We confirmed this agreement of the indices up to $O(x^3)$.

%%%%%%%%%%%%%%%%%%%%%%%%%%%%%%%%%%%%%%%%%%%%%%%%%%%%%%%%%%
\subsection{3d $SU(N)$ gauge theory with $(N+4) \,  \protect\Young[0]{1}+\overline{ \protect\Young[0]{2}}$ and its $Spin(7)$ dual}
%%%%%%%%%%%%%%%%%%%%%%%%%%%%%%%%%%%%%%%%%%%%%%%%%%%%%%%%%%
We can also construct an $SU(N)$ Seiberg duality with a matter in a symmetric representation by swapping roles of the electric and magnetic descriptions studied above. This can be easily done by introducing two singlets $M,Y^{dressed}$ and an additional superpotential $\Delta W = M M_{QQ} +Y^{dressed} M_{SS}$ in the magnetic theory (Table \ref{Spin7magnetic}), which lifts the two mesons and simplifies the superpotential. The low-energy limit becomes a 3d $\mathcal{N}=2$ $SU(N)$ gauge theory with $N+3$ fundamental matters $Q$, a fundamental matter $Q'$ and a symmetric-bar tensor $S$. For simplicity, we shifted the rank of the gauge group. The superpotential becomes
\begin{align}
W_{ele}=\bar{S}Q'Q',
\end{align}
where we used capital letters for labeling the elementary fields since the theory is now ``electric.'' The Higgs branch is described by the following operators
\begin{align}
M:=\bar{S}QQ,~~~B:=Q^N,~~~B':=Q^{N-1}Q',~~~U:= \det \bar{S}.
\end{align}
Notice that the composite $\det \, \bar{S}$ is not truncated since the superpotential is simplified. We next investigate the Coulomb branch:
When the bare Coulomb branch $Y_{SU(N-2)}^{bare}$ obtains a non-zero vev, the gauge group is spontaneously broken to $SU(N) \rightarrow SU(N-2) \times U(1)_1 \times U(1)_2$ and the branching rules of the matter fields become
\begin{align} 
{\tiny \yng(1)}  & \rightarrow  {\tiny \yng(1)}_{\, 0,-2} +\mathbf{1}_{1,N-2}+\mathbf{1}_{-1,N-2} \\
{\tiny \overline{\yng(2)}} & \rightarrow  {\tiny \overline{\yng(2)}}_{\,0,4}+ {\tiny \overline{\yng(1)}}_{\, -1,-N+4}+{\tiny \overline{\yng(1)}}_{\, 1,-N+4}+\mathbf{1}_{-2,-2N+4}+\mathbf{1}_{2,-2N+4}+\mathbf{1}_{0,-2N+4} \\
\mathbf{adj.} & \rightarrow  \mathbf{adj.}_{0,0}+\mathbf{1}_{0,0}+\mathbf{1}_{0,0}+   \mathbf{1}_{2,0}+\mathbf{1}_{-2,0} \nonumber \\&\qquad + {\tiny \yng(1)} _{\,-1,-N}+{\tiny \yng(1)}_{\, 1,-N}+{\tiny \overline{ \yng(1)}}_{\, 1,N}+{\tiny \overline{ \yng(1)}}_{\, -1,N},
\end{align}
where the Coulomb branch is associated with the unbroken $U(1)_1$ subgroup and its coordinate is obtained by dualizing the $U(1)_1$ vector superfield. The components charged under $U(1)_1$ are massive along the flat direction $Y_{SU(N-2)}^{bare}$. Since the matter content is ``chiral,'' the mixed Chern-Simons term between $U(1)_1$ and $U(1)_2$ is generated due to these massive components. Therefore, $Y_{SU(N-2)}^{bare}$ has a non-zero $U(1)_2$ charge \cite{Intriligator:2013lca}. We therefore need to define baryon-monopole operators  
\begin{align}
Y^{dressed} &:=Y_{SU(N-2)}^{bare} \left( {\tiny \overline{\yng(2)}}_{\,0,4} \right)^{N-2}     \sim Y_{SU(N-2)}^{bare} \bar{S}^{N-2} \\
Y^{dressed}_{QW}&:=  Y_{SU(N-2)}^{bare} ( {\tiny \overline{\yng(1)}}_{\,0,2}  )^{N-4} {\tiny \overline{ \yng(1)}}_{\, 1,N}{\tiny \overline{ \yng(1)}}_{\, -1,N}  \sim Y_{SU(N-2)}^{bare} (\bar{S}Q)^{N-4}W_\alpha^2 \\
Y^{dressed}_Q&:= Y_{SU(N-2)}^{bare}  ( {\tiny \overline{\yng(1)}}_{\,0,2} )^{N-2} ( {\tiny \overline{\yng(2)}}_{\,0,4}^{N-2} \mathbf{1}_{0,-2N+4})  \sim Y_{SU(N-2)}^{bare} (\bar{S}Q)^{N-2} (\bar{S}^{N-2} \bar{S}).
\end{align}
Note that these operators are not truncated by the superpotential.

\begin{table}[H]\caption{3d $\mathcal{N}=2$ $SU(N)$ gauge theory with $(N+4) \,  \protect\Young[0]{1}+\overline{ \protect\Young[0]{2}}$} 
\begin{center}
\scalebox{0.95}{
  \begin{tabular}{|c||c|c|c|c|c| } \hline
  &$SU(N)$&$SU(N+3)$&$U(1)$&$U(1)$& $U(1)_R$  \\ \hline
$Q$&${\tiny \yng(1)}$&${\tiny \yng(1)}$&$1$&$0$&$0$ \\  
$Q'$&${\tiny \yng(1)}$&1&$0$&$1$&$0$ \\  
 $\bar{S}$&${\tiny \overline{\yng(2)}}$&1&$0$&$-2$&$2$ \\   \hline
%$M_{QQ}$ &1&${\tiny \yng(2)}$&2&0&0  \\
%$M_{SS}$ &1&1&0&2&0  \\  \hline
%$B:=Q^7$&1&7-th anti-symm. &7&0&0  \\
$M:=\bar{S}QQ$&1&${\tiny \yng(2)}$&2&$-2$&2  \\
$B:=Q^N$&1&\scriptsize $\overline{\mbox{3rd anti-symm.}}$&$N$&0&0  \\
$B':=Q^{N-1}Q'$&1&\scriptsize $\overline{\mbox{4-th anti-symm.}}$&$N-1$&1&0 \\
$U:= \det \bar{S}$&1&1&0&$-2N$&$2N$  \\ \hline
\scriptsize $Y_{SU(N-2)}^{bare}$&\scriptsize $U(1)_2$: $-4(N-2)$&1&$-N-3$&$2N-1$&$-2N+6$ \\ 
\scriptsize $Y^{dressed}:= Y_{SU(N-2)}^{bare} \bar{S}^{N-2}$&1&1&$-N-3$&$3$&$2$  \\  
\scriptsize $Y^{dressed}_{QW}:= Y_{SU(N-2)}^{bare} (\bar{S}Q)^{N-4}W_\alpha W_\alpha$&1&\scriptsize $\overline{\mbox{7-th anti-symm.}}$&$-7$&$7$&$0$  \\
\scriptsize $Y^{dressed}_Q:= Y_{SU(N-2)}^{bare} (\bar{S}Q)^{N-2} (\bar{S}^{N-2} \bar{S})$&1&\scriptsize $\overline{\mbox{5-th anti-symm.}}$&$-5$&$5-2N$&$2N$  \\ \hline
  \end{tabular}}
  \end{center}\label{SUSpin7ele}
\end{table}

The magnetic description becomes a 3d $\mathcal{N}=2$ $Spin(7)$ gauge theory with $N+3$ vectors $q$, one spinor $s$ and two gauge-singlets $M$ and $Y^{dressed}$. Since the $Spin(7)$ theory is ``magnetic,'' the elementary fields are represented by lowercase letters. The tree-level superpotential becomes
\begin{align}
W_{mag}=Mqq+Y^{dressed} ss,
\end{align}
which lifts all the magnetic mesons $qq$ and $ss$. The charge assignment is completely fixed by the superpotential. The operator identification under the duality is straightforwardly followed from the previous analysis and it is summarized in Table \ref{SUSpin7mag}.

\begin{table}[H]\caption{The magnetic $Spin(7)$ gauge theory dual to Table \ref{SUSpin7ele}} 
\begin{center}
\scalebox{1}{
  \begin{tabular}{|c||c|c|c|c|c| } \hline
  &$Spin(7)$&$SU(N+3)$&$U(1)$&$U(1)$& $U(1)_R$  \\ \hline
$q$&$\mathbf{7}$&${\tiny \overline{\yng(1)}}$&$-1$&$1$&$0$ \\  
$s$&$\mathbf{8}$&1&$\frac{N+3}{2}$&$-\frac{3}{2}$&$0$ \\  
$M$&1&${\tiny \yng(2)}$&2&$-2$&2  \\
$Y^{dressed}$&1&1&$-N-3$&$3$&$2$  \\  \hline
%$M_{QQ}$ &1&${\tiny \yng(2)}$&2&0&0  \\
%$M_{SS}$ &1&1&0&2&0  \\  \hline
%$B:=Q^7$&1&7-th anti-symm. &7&0&0  \\
$B \sim sq^3s$&1&\scriptsize $\overline{\mbox{3rd anti-symm.}}$&$N$&0&0  \\
$B' \sim sq^4s$&1&\scriptsize $\overline{\mbox{4-th anti-symm.}}$&$N-1$&1&0 \\
$Y^{dressed}_{QW} \sim q^7$&1&\scriptsize $\overline{\mbox{7-th anti-symm.}}$&$-7$&$7$&$0$  \\ \hline
$U \sim \tilde{Y}_{SO(5)}$&1&1&0&$-2N$&$2N$ \\
$Y^{dressed}_Q \sim \tilde{Y}_{SO(5)} q^5$&1&\scriptsize $\overline{\mbox{5-th anti-symm.}}$&$-5$&$5-2N$&$2N$  \\ \hline
  \end{tabular}}
  \end{center}\label{SUSpin7mag}
\end{table}

%%%%%%%%%%%%%%%%%%%%%%%%%%%%%%%%%%%%%%%%%%%%%%%%%%%%%%%%%%
\subsection{3d $Spin(7)$ gauge theory with $F$ vectors and $2$ spinors \label{Spin72spinors}}
%%%%%%%%%%%%%%%%%%%%%%%%%%%%%%%%%%%%%%%%%%%%%%%%%%%%%%%%%%
In this subsection, we will consider the $Spin(7)$ duality with two spinor matters (see \cite{Cho:1997kr} for the corresponding 4d duality). The electric description is a 3d $\mathcal{N}=2$ $Spin(7)$ gauge theory with $F$ vectors $Q$ and two spinors $S$. The global symmetry now becomes $SU(F) \times SU(2) \times U(1) \times U(1) \times U(1)_R$, where the $SU(2)$ rotates two spinors. The quantum numbers of the matter fields are summarized in Table \ref{Spin72spelectric}. The Higgs branch operators are defined by
\begin{gather*}
M_{QQ}:=Q^2,~~M_{SS}:=S^2,~~P_{3S}:=SQ^3S,~~P_{4S}:=SQ^4S,~~B:=Q^7 \\
P_{1A}:=SQS,~~P_{2A}:=SQ^2S,~~ P_{5A}:=SQ^5S,~~P_{6A}:=SQ^6S
\end{gather*}
where the color indices of $S$'s in $P_{3S}, P_{4S}$ and $M_{SS}$ are symmetrized. On the second line, the two spinor fields are anti-symmetrized. We summarized the quantum numbers of these operators in Table \ref{Spin72spelectric}.

In this example, there are two types of Coulomb branch operators, which is different from the previous situation. The first Coulomb branch, denoted by $Y_{SO(5)}$, is the same as the previous one and corresponds to the breaking $so(7) \rightarrow so(5) \times u(1)$. Along the $Y_{SO(5)}$ direction, the vector matters are reduced to massless vectors of $so(5)$ while all the components of the spinor matters are massive. The vacuum of the low-energy $so(5)$ gauge theory can be made supersymmetric and stable by the massless components of the $F$ vectors. Since there are massless components of the vector matters, we can also define a dressed operator $Y_{SO(5)}^{Q}:=Y_{SO(5)}Q^5$, which is known as a baryon-monopole operator \cite{Aharony:2013kma}. 

In $Spin(7)$ gauge theories with more than one spinor, we need to take into account another Coulomb branch \cite{Aharony:2013kma, Nii:2018tnd}. We denote this flat direction by $Z_{SO(3)}$ whose expectation value corresponds to the gauge symmetry breaking
\begin{align}
so(7) & \rightarrow so(3) \times su(2) \times u(1)  \\
\mathbf{7} & \rightarrow (\mathbf{3},\mathbf{1})_0+(\mathbf{1},\mathbf{2})_{\pm 1} \\
\mathbf{8} & \rightarrow (\mathbf{2},\mathbf{2})_0+(\mathbf{2},\mathbf{1})_{\pm 1},
\end{align}
where $Z_{SO(3)}$ is associated with the $U(1)$ subgroup. 
Along the $Z_{SO(3)}$ direction, the vector matter reduces to a massless component $(\mathbf{3},\mathbf{1})_0$. The theory only with vector matters has an unstable (runaway) vacuum of the low-energy $SU(2)$ pure SYM theory \cite{Affleck:1982as, Aharony:1997bx, deBoer:1997kr}. On the other hand, the spinor matters reduce to $(\mathbf{2}, \mathbf{2})_0$ which can make the low-energy $so(3) \times su(2)$ vacuum stable and supersymmetric. For the theory with a single spinor, the origin of the moduli space of the low-energy $SU(2)$ theory is eliminated \cite{Aharony:1997bx}. Therefore, the theory with more than one spinor should have this flat direction. 
Similar to the first Coulomb branch, we can also define a baryon-monopole operator
\begin{align}
Z_{SO(3)}^{Q}:=Z_{SO(3)}  ((\mathbf{3},\mathbf{1})_0)^3   \sim Z_{SO(3)}Q^3,
\end{align}
where the color indices of $Q^3$ is contracted by an epsilon tensor of $so(3)$. This operator is available when $F \ge 3$.

For small flavors with $F \le 3$, we find confinement descriptions. From the symmetry argument in Table \ref{Spin72spelectric}, the effective superpotential becomes
\begin{gather*}
W_{F=3}^{eff} =Z_{SO(3)} \left[ M_{QQ}^3 M_{SS}^2+M_{QQ}^2 P_{1A}^2 +M_{QQ}P_{2A}^2+P_{3S}^2 \right]+Z_{SO(3)}^Q \left[ P_{1A}P_{2A}+M_{SS}P_{3S} \right]  \\
W_{F=2}^{eff} = \lambda \left[ Z_{SO(3)} (M_{QQ}^2 M_{SS}^2 +M_{QQ}P_{1A}^2+P_{2A}^2) -1 \right]  \\
W_{F=1}^{eff} = \frac{1}{Z_{SO(3)}(M_{QQ}M_{SS}^2 +P_{1A}^2)},~~~W_{F=0}^{eff} = \left(   \frac{1}{Z_{SO(3) M_{SS}^2} }  \right)^{\frac{1}{2}},
\end{gather*}
where $\lambda$ is a Lagrange multiplier field. For $F=3$, the theory exhibits an s-confinement phase as studied in \cite{Nii:2018tnd}. For $F=2$, the moduli space is described by $M_{QQ},M_{SS},P_{1A},P_{2A}$ and $Z_{SO(3)}$ with a single constraint. The constraint removes the origin of the moduli space and then some of the global symmetries are inevitably broken.
For $F \le 1$, the theory shows a runaway superpotential and there is no stable supersymmetric vacuum.

\begin{table}[H]\caption{3d $\mathcal{N}=2$ $Spin(7)$ gauge theory with $F$ vectors and $2$ spinors} 
\begin{center}
\scalebox{1}{
  \begin{tabular}{|c||c|c|c|c|c|c| } \hline
  &$Spin(7)$&$SU(F)$&$SU(2)$&$U(1)$&$U(1)$& $U(1)_R$  \\ \hline
$Q$&$\mathbf{7}$&${\tiny \yng(1)}$&1&1&0&$0$ \\   
 $S$&$\mathbf{8}$&1&${\tiny \yng(1)}$&0&1&$0$ \\  \hline 
$M_{QQ}:=Q^2$ &1&${\tiny \yng(2)}$&1&2&0&0  \\
$M_{SS}:=S^2$ &1&1&${\tiny \yng(2)}$&0&2&0  \\
$P_{1A}:=SQS$&1&${\tiny \yng(1)}$&1&1&2&0  \\
$P_{2A}:=SQ^2S$&1&${\tiny \yng(1,1)}$&1&2&2&0  \\[4pt]
$P_{3S}:=SQ^3S$&1&${\tiny \yng(1,1,1)}$&${\tiny \yng(2)}$&3&2&0  \\
$P_{4S}:=SQ^4S$&1&\scriptsize 4-th anti-symm.&${\tiny \yng(2)}$&4&2&0  \\  
$P_{5A}:=SQ^5S$ &1&\scriptsize 5-th anti-symm.&1&5&2&0  \\ 
$P_{6A}:=SQ^6S$ &1&\scriptsize 6-th anti-symm.&1&6&2&0  \\  
$B:= Q^7$&1&\scriptsize 7-th anti-symm.&1&7&0&0  \\ \hline
$Y_{SO(5)}$&1&1&1&$-2F$&$-8$&$2F-2$  \\
$Z_{SO(3)}$&1&1&1&$-2F$&$-4$&$2F-4$  \\
$Y_{SO(5)}^{Q}:=Y_{SO(5)}Q^5$&1&\scriptsize 5-th anti-symm.&1&$5-2F$&$-8$&$2F-2$  \\
$Z_{SO(3)}^{Q}:=Z_{SO(3)}Q^3$&1&${\tiny \yng(1,1,1)}$&1&$3-2F$&$-4$&$2F-4$  \\[5pt] \hline
  \end{tabular}}
  \end{center}\label{Spin72spelectric}
\end{table}

%%%%%%%%%%%%%%%%%%%%%%%%%%%%%%%%%%%%%%%%%%%%%%%%%%%%%%%%%
The magnetic description is given by a 3d $\mathcal{N}=2$ $SU(F-2)$ gauge theory with $F+3$ fundamental matters, an anti-fundamental matter and a symmetric-bar tensor. The theory includes the mesons $M_{QQ}, M_{SS}$ and $ P_{1A}$ as elementary fields, which are easily identified with the electric mesons in Table \ref{Spin72spelectric}. The magnetic theory has a tree-level superpotential
\begin{align}
W_{mag}= M_{QQ} \bar{s}qq +\bar{s}q'q' +P_{1A} q \bar{q} +M_{SS} q' \bar{q} + \tilde{Y}^{dressed}, \label{Spin72spmagW}
\end{align}
which distinguishes three fundamental quarks $q'$ from the other quarks $q$. Therefore, the global symmetry becomes $SU(F) \times SO(3) \times U(1) \times U(1) \times U(1)_R$ which is isomorphic to the global symmetry on the electric side\footnote{The gauge-invariant operators defined above and below do not distinguish the difference between $SU(2)$ and $SO(3)$. Therefore, the (faithful) global symmetry becomes $SO(3)$ in the far-infrared limit.}. 
Notice that the charge assignment of the magnetic elementary fields is completely determined by the above superpotential and hence the matching of gauge-invariant operators becomes a non-trivial consistency check of the duality. The last term in \eqref{Spin72spmagW} is a dressed Coulomb branch operator which will be defined below. The matching of the Higgs branch operators are obtained as follows:
\begin{gather*}
P_{2A} \sim q^{F-2},~~~P_{3S} \sim q^{F-3}q',~~~P_{4S} \sim q^{F-4}q'^2,~~~P_{5A} \sim q^{F-5} q'^3 \\
Z_{SO(3)} \sim \bar{s}^{F-2},~~~Y_{SO(5)} \sim \bar{s}^{F-3} \bar{q}^2,~~~Z_{SO(3)}^{Q} \sim (\bar{s}q)^{F-3} \bar{q},
\end{gather*}
where the color indices are contracted by epsilon tensors of $SU(F-2)$. The mesonic operators on the magnetic side are all lifted by the superpotential.

When the bare Coulomb branch operator $\tilde{Y}_{SU(F-4)}^{bare}$ obtains a non-zero expectation value, the gauge symmetry is spontaneously broken to
\begin{align}
SU(F-2) & \rightarrow SU(F-4) \times U(1)_1 \times U(1)_2 \\ 
{\tiny \yng(1)}  & \rightarrow  {\tiny \yng(1)}_{\, 0,-2} +\mathbf{1}_{1,F-4}+\mathbf{1}_{-1,F-4}\\
{\tiny \overline{\yng(1)}}  & \rightarrow  {\tiny \overline{\yng(1)}}_{\, 0,2} +\mathbf{1}_{-1,-(F-4)}+\mathbf{1}_{1,-(F-4)}\\
{\tiny \overline{ \yng(2)}} & \rightarrow {\tiny \overline{ \yng(2)}}_{\, 0,4} +{\tiny \overline{ \yng(1)}}_{\,-1,-F+6}  +{\tiny \overline{ \yng(1)}}_{\, 1,-F+6} +\mathbf{1}_{-2,-2F+8} +\mathbf{1}_{2,-2F+8}+\mathbf{1}_{0,-2F+8}   \\
\mathbf{adj.} & \rightarrow  \mathbf{adj.}_{0,0}+\mathbf{1}_{0,0}+\mathbf{1}_{0,0}+   \mathbf{1}_{2,0}+\mathbf{1}_{-2,0} \nonumber \\&\qquad + {\tiny \yng(1)} _{\,-1,-F+2}+{\tiny \yng(1)}_{\, 1,-F+2}+{\tiny \overline{ \yng(1)}}_{\, 1,F-2}+{\tiny \overline{ \yng(1)}}_{\, -1,F-2},
\end{align}
where the Coulomb branch corresponds to the $U(1)_1$ generator. Along the Coulomb branch $\tilde{Y}_{SU(F-4)}^{bare}$, the components charged under the $U(1)_1$ symmetry become massive and integrated out. This results in a mixed Chern-Simons term between $U(1)_1$ and $U(1)_2$, which induces a $U(1)_2$ charge of $\tilde{Y}_{SU(F-4)}^{bare}$ as in Table \ref{Spin72spmagnetic} .  
Therefore, we need to construct baryon-monopoles for defining gauge-invariant operators. The dressed Coulomb branch operators are defined as follows:
\begin{align}
\tilde{Y}^{dressed} &:=\tilde{Y}_{SU(F-4)}^{bare} \left( {\tiny \overline{ \yng(2)}}_{\, 0,4}  \right)^{F-4}   \sim \tilde{Y}_{SU(F-4)}^{bare} \bar{s}^{F-4}\\
P_{6A} &:=\tilde{Y}_{SU(F-4)}^{bare} ({\tiny  \overline{\yng(1)}}_{\, 0,2})^{F-6} {\tiny \overline{ \yng(1)}}_{\, 1,F-2}{\tiny \overline{ \yng(1)}}_{\, -1,F-2}   \nonumber \\ 
& \sim  \tilde{Y}_{SU(F-4)}^{bare} (\bar{s}q)^{F-6} \tilde{W}_\alpha^2 \\
B&:= \tilde{Y}_{SU(F-4)}^{bare} ({\tiny  \overline{\yng(1)}}_{\, 0,2})^{F-7}{\tiny  \overline{\yng(1)}}_{\, 0,2} {\tiny \overline{ \yng(1)}}_{\, 1,F-2}{\tiny \overline{ \yng(1)}}_{\, -1,F-2}   \nonumber \\ 
& \sim  \tilde{Y}_{SU(F-4)}^{bare} (\bar{s}q)^{F-7} \bar{q} \tilde{W}_\alpha^2 \\
Y_{SO(5)}^Q &:=  \tilde{Y}_{SU(F-4)}^{bare} \left( {\tiny \overline{ \yng(2)}}_{\, 0,4}  \right)^{F-5} {\tiny \overline{ \yng(2)}}_{\, 0,4}  \mathbf{1}_{0,-2F+8} ({\tiny  \overline{\yng(1)}}_{\, 0,2})^{F-5}   {\tiny  \overline{\yng(1)}}_{\, 0,2}   \nonumber \\
& \sim \tilde{Y}_{SU(F-4)}^{bare} \bar{s}^{F-5} \bar{q}^2 \bar{s} (\bar{s}q)^{F-5} \bar{q},
\end{align}
where the color indices of the matter chiral superfields are contracted by epsilon tensors of the unbroken $SU(F-4)$ subgroup. 
Notice that we can't use $\bar{s}q'$ for constructing gauge invariant baryon-monopoles because of the $F$-term condition for $q'$. The first operator $\tilde{Y}^{dressed} $ is eliminated from the chiral ring elements due to the superpotential. The other operators are correctly mapped to the electric operators as indicated in Table \ref{Spin72spmagnetic}.

\begin{table}[H]\caption{The magnetic description dual to Table \ref{Spin72spelectric}} 
\begin{center}
\scalebox{0.85}{
  \begin{tabular}{|c||c|c|c|c|c|c| } \hline
  &$SU(F-2)$&$SU(F)$&$SO(3)$&$U(1)$&$U(1)$& $U(1)_R$  \\ \hline
$q$ &${\tiny \yng(1)}$&${\tiny \overline{\yng(1)}}$&1&$\frac{2}{F-2}$&$\frac{2}{F-2}$&0  \\
$q'$ &${\tiny \yng(1)}$&1&${\tiny \yng(1)}$&$\frac{F}{F-2}$&$\frac{2}{F-2}$&0   \\
$\bar{q}$&${\tiny \overline{\yng(1)}}$&1&1&$-\frac{F}{F-2}$&\footnotesize  $-2-\frac{2}{F-2}$&2  \\
$\bar{s}$ &${\tiny \overline{\yng(2)}}$&1&1&$-\frac{2F}{F-2}$&$-\frac{4}{F-2}$&2  \\
$M_{QQ}$ &1&${\tiny \yng(2)}$&1&2&0&0  \\
$M_{SS}$ &1&1&${\tiny \yng(1)}$&0&2&0  \\
$P_{1A}$&1&${\tiny \yng(1)}$&1&1&2&0  \\ \hline
$P_{2A} \sim q^{F-2}$&1&${\tiny \yng(1,1)}$&1&2&2&0  \\[3pt]
$P_{3S} \sim q^{F-3}q'$&1&${\tiny \yng(1,1,1)}$&${\tiny \yng(1)}$&3&2&0  \\[5pt]
$P_{4S} \sim q^{F-4}q'^2$&1&\scriptsize 4-th anti-symm.&${\tiny \yng(1)}$&4&2&0  \\
$P_{5A} \sim q^{F-5} q'^3$&1&\scriptsize 5-th anti-symm.&1&5&2&0 \\
$Z_{SO(3)} \sim \bar{s}^{F-2}$&1&1&1&$-2F$&$-4$&$2F-4$   \\
$Y_{SO(5)} \sim \bar{s}^{F-3} \bar{q}^2$&1&1&1&$-2F$&$-8$&$2F-2$   \\ 
$Z_{SO(3)}^{Q} \sim (\bar{s}q)^{F-3} \bar{q}$ &1&${\tiny \yng(1,1,1)}$&${\tiny \yng(1)}$&$3-2F$&$-4$& $2F-4$  \\[4pt]  \hline
\footnotesize $\tilde{Y}_{SU(F-4)}^{bare} $&\scriptsize $U(1)_2$: $-4(F-4)$&1&1& \scriptsize $2F-4-\frac{8}{F-2}$&\scriptsize$4-\frac{8}{F-2}$&\footnotesize  $-2F+10$   \\
\footnotesize $\tilde{Y}^{dressed}:=\tilde{Y}_{SU(F-4)}^{bare} \bar{s}^{F-4} $&1&1&1&0&0&2  \\
\footnotesize $P_{6A} \sim \tilde{Y}_{SU(F-4)}^{bare} (\bar{s}q)^{F-6} \tilde{W}_\alpha^2$&1&\scriptsize 6-th anti-symm.&1&6&2&0  \\
\footnotesize $B\sim  \tilde{Y}_{SU(F-4)}^{bare} (\bar{s}q)^{F-7} \bar{q} \tilde{W}_\alpha^2$&1&\scriptsize 7-th anti-symm.&1&7&0&0  \\
\scriptsize $Y_{SO(5)}^Q \sim \tilde{Y}_{SU(F-4)}^{bare} \bar{s}^{F-5} \bar{q}^2 \bar{s} (\bar{s}q)^{F-5} \bar{q}  $&1&\scriptsize 5-th anti-symm.&1&$5-2F$&$-8$&$2F-2$  \\ \hline
  \end{tabular}}
  \end{center}\label{Spin72spmagnetic}
\end{table}

Finally, we will test the superconformal indices for the $F=4$ case which is simple enough to explicitly compute since the dual gauge group becomes $SU(2)$. On both the electric and magnetic sides, we obtained the following result:
\footnotesize
\begin{align}
I_{F=4} &= 1+x^{1/2} \left(10 t^2+3 u^2\right)+4 t u^2 x^{3/4}+x \left(\frac{1}{t^8 u^4}+55 t^4+36 t^2 u^2+6 u^4\right) \nonumber  \\
&+x^{5/4} \left(52 t^3 u^2+12 t u^4\right)+x^{3/2} \left(\frac{3}{t^8 u^2}+\frac{10}{t^6 u^4}+220 t^6+228 t^4 u^2+88 t^2 u^4+10 u^6\right) \nonumber \\
&+x^{7/4} \left(\frac{4}{t^7 u^2}+\frac{4}{t^5 u^4}+340 t^5 u^2+176 t^3 u^4+24 t u^6\right)  \nonumber \\
&+x^2 \left(\frac{1}{t^{16} u^8}+\frac{1}{t^8 u^8}+715 t^8+\frac{6}{t^8}+1020 t^6 u^2+\frac{36}{t^6 u^2}+684 t^4 u^4+\frac{55}{t^4 u^4}+166 t^2 u^6+15 u^8-20\right)+\cdots,
\end{align}
\normalsize

\noindent where $t$ and $u$ are the fugacities for the vector and spinor $U(1)$ symmetries. The r-charges are set to be $r_{Q}=r_{S} =\frac{1}{4}$ for simplicity. One can easily check the agreement of the indices for other r-charges. We verified the agreement up to $O(x^2)$. The second term $x^{1/2} \left(10 t^2+3 u^2\right)$ corresponds to the two mesons $M_{QQ}+M_{SS}$. The third term $4 t u^2 x^{3/4}$ is regarded as $P_{1A}$. The fourth term $x \left(\frac{1}{t^8 u^4}+55 t^4+36 t^2 u^2+6 u^4\right)$ includes $Z_{SO(3)}+M_{QQ}^2+M_{QQ}M_{SS}+P_{2A}+M_{SS}^2$, where $M_{QQ}^2$ and $M_{SS}^2$ are symmetric products. The fifth term $x^{5/4} \left(52 t^3 u^2+12 t u^4\right)$ is identified with $M_{QQ}P_{1A}+P_{3S}+M_{SS}P_{1A}$. The remaining operators $P_{4S}, Y_{SO(5)} $ and $ Z_{SO(3)}^Q$ are represented as $3 t^4 u^2 x^{3/2}, \frac{x^2}{t^8 u^8}$ and $\frac{x^{7/4}}{t^5u^4}$, respectively.

%%%%%%%%%%%%%%%%%%%%%%%%%%%%%%%%%%%%%%%%%%%%%%%%%%%%%%%%%%
%%%%%%%%%%%%%%%%%%%%%%%%%%%%%%%%%%%%%%%%%%%%%%%%%%%%%%%%%%
\section{3d $Spin(8)$ Seiberg duality}
%%%%%%%%%%%%%%%%%%%%%%%%%%%%%%%%%%%%%%%%%%%%%%%%%%%%%%%%%%
%%%%%%%%%%%%%%%%%%%%%%%%%%%%%%%%%%%%%%%%%%%%%%%%%%%%%%%%%%
In this section, we will study the Seiberg duality for the $Spin(8)$ gauge group with spinor and vector matters. The corresponding 4d dualities were investigated in \cite{Pouliot:1995sk, Cho:1997kr}. Here, we propose its 3d version.
The electric description is a 3d $\mathcal{N}=2$ $Spin(8)$ gauge theory with $F$ vectors $Q$ and a spinor $S$. The spinor is denoted as $\mathbf{8_s}$. This theory is completely the same as the 4d theory except for the fact that the global symmetry is enhanced due to the absence of chiral anomalies. Table \ref{Spin8electric} summarizes the charge assignment of the elementary fields under the local and global symmetries.
The Higgs branch is described by the following gauge-invariant operators
\begin{align}
M_{QQ}:=Q^2,~~~M_{SS}:=S^2,~~~P_4:=SQ^4S,~~~B:=Q^8,
\end{align}
where $P_4$ is available for $F \ge 4$ and $B$ is defined only for $F \ge 8$. The flavor indices of $P_4$ and $B$ are anti-symmetrized. The mesons $M_{QQ}$ and $M_{SS}$ will become elementary fields in a magnetic theory.

Next, we study the Coulomb moduli space of the $Spin(8)$ gauge theory. For the description of the s-confinement (the $F=5$ case), the Coulomb branch was studied in \cite{Nii:2018wwj} (see also \cite{Aharony:2011ci, Aharony:2013kma}). When the Coulomb branch operator $Y_{SO(6)}$ obtains an expectation value, the gauge group is spontaneously broken as
\begin{align}
so(8) & \rightarrow so(6) \times u(1) \\
\mathbf{8_v} & \rightarrow \mathbf{6}_0 +\mathbf{1}_2 +\mathbf{1}_{-2} \\
\mathbf{8_s} & \rightarrow  \mathbf{4}_1 +\overline{\mathbf{4}}_{-1}  \\
\mathbf{8_c} & \rightarrow  \mathbf{4}_{-1} +\overline{\mathbf{4}}_{1},  
\end{align}
where we also listed a branching rule for a conjugate spinor $\mathbf{8_c}$ representation for the sake of Section \ref{Spin8F11}. The Coulomb branch $Y_{SO(6)}$ is associated with the unbroken $U(1)$ subgroup. 
All the components of the (conjugate) spinor representations are massive along the $Y_{SO(6)}$ Coulomb branch and do not contribute to the far-infrared physics. As a result, for theories only with (conjugate) spinor matters, the $Y_{SO(6)}$ branch obtains a runaway potential $W=\sum_{i=1,2,3} \frac{1}{Y_i}$ from the fundamental monopoles of $U(1)^3 \subset Spin(6)$ \cite{Affleck:1982as, Aharony:1997bx, deBoer:1997kr}. Therefore, $Y_{SO(6)}$ is eliminated from the moduli space for theories with only spinor matters. On the other hand, a vector representation reduces to a massless vector of the unbroken $SO(6)$ gauge dynamics. Therefore, the vacuum of the low-energy $SO(6)$ gauge theory can be made supersymmetric and stable (no runaway potential generated). In addition to this bare operator $Y_{SO(6)}$, we can also define a baryon-monopole operator
\begin{align}
Y_{SO(6)}^Q := Y_{SO(6)}  \left(  \mathbf{6}_0   \right)^6 \sim Y_{SO(6)}  Q^6, 
\end{align}
where the color indices of $Q^6$ are contracted by an epsilon tensor of $SO(6)$. $Y_{SO(6)}^Q$ is available for $F \ge 6$. The quantum numbers of the Higgs and Coulomb branch operators are summarized in Table \ref{Spin8electric}.

For small flavors $F \le 5$, we find confinement phases where the low-energy dynamics is described by the effective superpotential
\begin{align}
W_{F=5}^{eff} &= Y_{SO(6)}  \left(M_{QQ}^5 M_{SS}^2 +M_{QQ}P_4^2  \right) \\
W_{F=4}^{eff} &= \lambda   \left[ Y_{SO(6)} \left( M_{QQ}^4 M_{SS}^2+P_4^2 \right) -1  \right]  \\
W_{F\le 3}^{eff} &= \left( \frac{1}{Y_{SO(6)}  M^{F}_{QQ} M_{SS}^2 }   \right)^{\frac{1}{4-F}},
\end{align}
where $\lambda$ is a Lagrange multiplier field.
For $F=5$, the theory exhibits s-confinement as studied in \cite{Nii:2018wwj}. The low-energy dynamics is described by $M_{QQ}, M_{SS}, P_4$ and $Y_{SO(6)}$ and there is no singularity at the origin of moduli space. For $F=4$, these operators have one quantum constraint which removes the origin of the moduli space and we can remove one coordinate. For $F \le 3$, the effective superpotential becomes runaway and there is no stable supersymmetric vacuum.

\begin{table}[H]\caption{3d $\mathcal{N}=2$ $Spin(8)$ gauge theory with $F$ vectors and a spinor} 
\begin{center}
\scalebox{1}{
  \begin{tabular}{|c||c|c|c|c|c| } \hline
  &$Spin(8)$&$SU(F)$&$U(1)$&$U(1)$& $U(1)_R$  \\ \hline
$Q$&$\mathbf{8}_v$&${\tiny \yng(1)}$&1&0&$0$ \\   
 $S$&$\mathbf{8}_s$&1&0&1&$0$ \\  \hline 
$M_{QQ}:=Q^2$ &1&${\tiny \yng(2)}$&2&0&0  \\
$M_{SS}:=S^2$ &1&1&0&2&0  \\
%$B:=Q^7$&1&7-th anti-symm. &7&0&0  \\
%$P_3:=SQ^3S$ &1&${\tiny \yng(1,1,1)}$&3&2&0  \\[5pt]
$P_4:=SQ^4S$ &1&\scriptsize 4-th anti-symm.&4&2&0 \\   
$B:=Q^8 $&1&\scriptsize 8-th anti-symm.&8&0&0  \\ \hline
$Y_{SO(6)}$&1&1&$-2F$&$-4$&$2F-8$ \\ 
$Y_{SO(6)}^Q:=Y_{SO(6)}Q^6$&1&\scriptsize 6-th anti-symm.&$6-2F$&$-4$&$2F-8$ \\  \hline
  \end{tabular}}
  \end{center}\label{Spin8electric}
\end{table}

%%%%%%%%%%%%%%%%%%%%%%%%%%%%%%%%%%%%%%%%%%%%%%%%%%%%%%%%%%
Next, we move on to the magnetic description which is given by a 3d $\mathcal{N}=2$ $SU(F-4)$ gauge theory with $F$ fundamental matters $q$, a symmetric-bar tensor $\bar{s}$ and two gauge-singlet mesons $M_{QQ}$ and $M_{SS}$. These two singlets are straightforwardly identified with the electric counterparts. The magnetic theory has a tree-level superpotential
\begin{align}
W_{mag}=M_{QQ} \bar{s}qq +M_{SS} \tilde{Y}^{dressed}.
\end{align}
Note that the last term in the superpotential is again slightly different from the 4d one and that this difference is important to find a correct operator mapping which will be discussed in what follows.
The definition of the dressed Coulomb branch $\tilde{Y}^{dressed}$ will be given below. The consistency of the above superpotential completely fixes the charge assignment of the magnetic elementary fields as in Table \ref{Spin8magnetic}. By comparing the gauge-invariant composites in Table \ref{Spin8electric} and Table \ref{Spin8magnetic}, we can easily find the following matching 
\begin{align}
Y_{SO(6)} \sim \det \bar{s},~~~P_4 \sim q^{F-4}.
\end{align}
Notice that $(\bar{s}q)^{F-4}$ cannot be an independent operator.

As in the previous section, we can similarly study the Coulomb branch coordinates by carefully flowing to the semi-classical region of the Coulomb branch. When the bare Coulomb branch $\tilde{Y}^{bare}_{SU(F-6)}$ obtains a (large) non-zero vev, the gauge group is spontaneously broken to
\begin{align}
SU(F-4) & \rightarrow SU(F-6) \times U(1)_1 \times U(1)_2 \\ 
{\tiny \yng(1)}  & \rightarrow  {\tiny \yng(1)}_{\, 0,-2} +\mathbf{1}_{1,F-6}+\mathbf{1}_{-1,F-6}\\
{\tiny \overline{ \yng(2)}} & \rightarrow {\tiny \overline{ \yng(2)}}_{\, 0,4} +{\tiny \overline{ \yng(1)}}_{\,-1,-F+8}  +{\tiny \overline{ \yng(1)}}_{\, 1,-F+8} +\mathbf{1}_{-2,-2F+12} +\mathbf{1}_{2,-2F+12}+\mathbf{1}_{0,-2F+12}   \\
\mathbf{adj.} & \rightarrow  \mathbf{adj.}_{0,0}+\mathbf{1}_{0,0}+\mathbf{1}_{0,0}+   \mathbf{1}_{2,0}+\mathbf{1}_{-2,0} \nonumber \\&\qquad + {\tiny \yng(1)} _{\,-1,-F+4}+{\tiny \yng(1)}_{\, 1,-F+4}+{\tiny \overline{ \yng(1)}}_{\, 1,F-4}+{\tiny \overline{ \yng(1)}}_{\, -1,F-4},
\end{align}
where $\tilde{Y}^{bare}_{SU(F-6)}$ is associated with the unbroken $U(1)_1$ subgroup. By following the same argument as the previous section, the dressed gauge-invariant operators are defined as
\begin{align}
\tilde{Y}^{dressed} &:= \tilde{Y}^{bare}_{SU(F-6)}  ({\tiny \overline{ \yng(2)}}_{\, 0,4} )^{F-6} \sim  \tilde{Y}^{bare}_{SU(F-6)} \bar{s}^{F-6}   \\
B &:=  \tilde{Y}^{bare}_{SU(F-6)}  ({\tiny \overline{ \yng(1)}}_{\, 0,2})^{F-8} {\tiny \overline{ \yng(1)}}_{\, 1,F-4}{\tiny \overline{ \yng(1)}}_{\, -1,F-4} \sim \tilde{Y}^{bare}_{SU(F-6)} (\bar{s}q)^{F-8} \tilde{W}_\alpha \tilde{W}_\alpha \\
Y_{SO(6),Q} &:= \tilde{Y}^{bare}_{SU(F-6)}   ({\tiny \overline{ \yng(1)}}_{\, 0,2})^{F-6} ({\tiny \overline{ \yng(2)}}_{\, 0,4})^{F-6} \mathbf{1}_{0,-2F+12}  \sim  \tilde{Y}^{bare}_{SU(F-6)} (\bar{s}q)^{F-6} (\bar{s}^{F-6}\bar{s}),
\end{align}
where the color indices of the matter chiral superfields are contracted by epsilon tensors of the $SU(F-6)$ gauge group. 

\begin{table}[H]\caption{The magnetic description dual to Table \ref{Spin8electric}} 
\begin{center}
\scalebox{1}{
  \begin{tabular}{|c||c|c|c|c|c| } \hline
  &$SU(F-4)$&$SU(F)$&$U(1)$&$U(1)$& $U(1)_R$  \\ \hline
$q$&${\tiny \yng(1)}$&${\tiny \overline{\yng(1)}}$&$\frac{4}{F-4}$&$\frac{2}{F-4}$&$0$ \\  
%$q'$&${\tiny \yng(1)}$&1&$\frac{F}{F-3}$&$\frac{2}{F-3}$&$0$ \\  
 $\bar{s}$&${\tiny \overline{\yng(2)}}$&1&$\frac{-2F}{F-4}$&$\frac{-4}{F-4}$&$2$ \\  
$M_{QQ}$ &1&${\tiny \yng(2)}$&2&0&0  \\
$M_{SS}$ &1&1&0&2&0  \\  \hline
%$B:=Q^7$&1&7-th anti-symm. &7&0&0  \\
$Y_{SO(6)} \sim \det \bar{s}$&1&1&$-2F$&$-4$&$2F-6$  \\
$P_4 \sim q^{F-4}$ &1&\scriptsize 4-th anti-symm.&4&2&0  \\  \hline
%$P_4 \sim q^{F-4}q'$ &1&${\tiny \yng(1,1,1,1)}$&4&2&0 \\[7pt]   \hline
 \footnotesize $\tilde{Y}_{SU(F-6)}^{bare}$&\scriptsize $U(1)_2$: $-4(F-6)$&1&\footnotesize $2F-\frac{4F}{F-4}$& \footnotesize $4-\frac{2F}{F-4}$& \footnotesize$-2F+14$ \\ 
\scriptsize $\tilde{Y}^{dressed}:= \tilde{Y}_{SU(F-6)}^{bare} \bar{s}^{F-6}$&1&1&0&$-2$&2  \\  
\scriptsize $Y_{SO(6)}^Q \sim  \tilde{Y}_{SU(F-6)}^{bare} \bar{s}^{F-6} \bar{s}(\bar{s}q)^{F-6}$&1&\scriptsize 6-th anti-symm.& \footnotesize $6-2F$&$-4$&\footnotesize $2F-8$  \\
\scriptsize $B \sim  \tilde{Y}_{SU(F-6)}^{bare}  (\bar{s}q)^{F-8} \tilde{W}_\alpha^2$&1&\scriptsize 8-th anti-symm.&8&0&0  \\  \hline
  \end{tabular}}
  \end{center}\label{Spin8magnetic}
\end{table}

%%%%%%%%%%%%%%%%%%%%%%%%%%%%%%%%%%%%%%%%%%%%%%%%%%%%%%%%%%
We can study various consistency checks of the duality between Table \ref{Spin8electric} and Table \ref{Spin8magnetic}. First, we introduce a non-zero vev to $M_{SS}$ and study its induced flow. On the electric side, the gauge group is higgsed into $Spin(7)$ and the vector matters are transformed into $Spin(7)$ spinors. The spinor matter is eaten via the Higgs mechanism and then we obtain a 3d $\mathcal{N}=2$ $Spin(7)$ gauge theory with $F$ spinors. On the magnetic side, the vev for $M_{SS}$ brings the superpotential to the following form
\begin{align}
W_{mag}= M_{QQ} \bar{s}qq + \tilde{Y}^{dressed},
\end{align}
where we rescaled the dressed Coulomb branch and absorbed the vev $\braket{M_{SS}}$. This flow correctly reproduces the 3d $Spin(7)$ duality which was studied in \cite{Nii:2019wjz}. 

When $\braket{M_{QQ}}$ obtains a rank-1 vacuum expectation value, the electric gauge group is higgsed to $Spin(7)$ and one vector matter is eaten. On the magnetic side, the fundamental matters are decomposed into $F-1$ quarks $q$ and a single quark $q'$. The magnetic superpotential is decomposed into 
\begin{align}
W_{mag} =M_{QQ} \bar{s}qq +\bar{s}q'q' + M_{SS} \tilde{Y}^{dressed}
\end{align}
and thus we reproduce the duality studied in the previous section.

Next, we consider adding a complex mass to a single vector matter, let's say $W=mQ^FQ^{F}=m M^{FF}_{QQ}$. On the electric side, the vector matter $Q^F$ is integrated out, which results in a shift $F \rightarrow F-1$. On the magnetic side, $W=m M_{QQ}^{FF}$ breaks the gauge group into $SU(F-5)$ and reduces the number of fundamental quarks by one. Therefore, the duality is correctly preserved via a complex mass $W=mQQ=m M_{QQ}$ with a shift of $F$.

Alternatively, we can also introduce a complex mass as $\Delta W=m SS=m M_{SS}$. On the electric side, the spinor matter is integrated out and the low-energy limit becomes a 3d $\mathcal{N}=2$ $Spin(8)$ gauge theory with $F$ vectors. On the magnetic side, the mass term corresponds to the higgsing $\braket{\tilde{Y}^{dressed}} =-m$. Since $\tilde{Y}^{dressed}$ is a composite of $\tilde{Y}_{SU(F-6)}^{bare} $ and $\bar{s}^{F-6}$, the magnetic gauge group is broken as
\begin{align}
SU(F-4) & \rightarrow S(U(F-6) \times U(1)_1 \times U(1)_2 ) \\
 &\rightarrow S(O(F-6) \times U(1)_1  )  \cong  S(O(F-6) \times O(2) ),
\end{align}
where the first breaking is induced by $Y_{SU(F-6)}^{bare}$ and the second breaking is from $\bar{s}^{F-6}$.  There is no dynamical quark under the $O(2)$ gauge group and then the $O(2)$ part can be dualized to a single chiral superfield $Y_{O(2)}^{-}$ which is charged under the charge conjugation symmetry of $ S(O(F-6) \times O(2) )$. The low-energy theory can be regarded as a 3d $\mathcal{N}=2$ $O(F-6)$ gauge theory with $F$ vector and $Y_{O(2)}^{-}$, where $Y_{O(2)}^{-}$ is odd under $\mathbb{Z}_2 \subset O(F-6)$. The magnetic superpotential becomes 
\begin{align}
W_{mag} =M_{QQ}qq +Y_{O(2)}^{-} \tilde{Y}_{O(F-6)}^{-}, 
\end{align}
where the second term is a Affleck-Harvey-Witten type superpotential generated via the breaking $SU(F-4)  \rightarrow S(O(F-6) \times O(2) )$. $\tilde{Y}_{O(F-6)}^{-}$ is a Coulomb branch operator of $O(F-6)$. This duality was studied in \cite{Aharony:2013kma} and verifies consistency of our duality.

%%%%%%%%%%%%%%%%%%%%%%%%%%%%%%%%%%%%%%%%%%%%%%%%%%%%%%%%%%
\subsection{3d $SU(N)$ gauge theory with $(N+4) \,  \protect\Young[0]{1}+\overline{ \protect\Young[0]{2}}$}
%%%%%%%%%%%%%%%%%%%%%%%%%%%%%%%%%%%%%%%%%%%%%%%%%%%%%%%%%%
By swapping the electric and magnetic descriptions, we can construct a 3d duality between $SU(N)$ and $Spin(8)$ gauge groups with a symmetric tensor. This duality can be easily obtained from the previous one by introducing two gauge singlets ($M$ and $Y^{dressed}$) and adding a flipping superpotential \eqref{Spin(8)swap} to the electric $Spin(8)$ gauge theory in Table \ref{Spin8electric}. The superpotential lifts the vector and spinor mesons in the $Spin(8)$ theory. By adding the same superpotential on the magnetic side and by integrating out the massive modes, we obtain the following duality.

The electric description is a 3d $\mathcal{N}=2$ $SU(N)$ gauge theory with $N+4$ fundamental matters $Q$ and a symmetric-bar tensor $\bar{S}$, where we replaced the size of the gauge group from $F-4$ to $N$ for simplicity. We used uppercase letters for representing the elementary fields since this theory is now ``electric.'' Since there is no superpotential, the Higgs branch operators are not truncated and described by 
\begin{align}
M:=\bar{S}QQ,~~~B:=Q^N,~~~U:= \det \bar{S}.
\end{align}

As with the previous analysis, the bare Coulomb branch, denoted by $Y_{SU(N-2)}^{bare}$, corresponds to the breaking  
\begin{align}
SU(N) & \rightarrow SU(N-2) \times U(1)_1 \times U(1)_2 \\
{\tiny \yng(1)}  & \rightarrow  {\tiny \yng(1)}_{\, 0,-2} +\mathbf{1}_{1,N-2}+\mathbf{1}_{-1,N-2} \\
{\tiny \overline{\yng(2)}} & \rightarrow  {\tiny \overline{\yng(2)}}_{\,0,4}+ {\tiny \overline{\yng(1)}}_{\, -1,-N+4}+{\tiny \overline{\yng(1)}}_{\, 1,-N+4}+\mathbf{1}_{-2,-2N+4}+\mathbf{1}_{2,-2N+4}+\mathbf{1}_{0,-2N+4} \\
\mathbf{adj.} & \rightarrow  \mathbf{adj.}_{0,0}+\mathbf{1}_{0,0}+\mathbf{1}_{0,0}+   \mathbf{1}_{2,0}+\mathbf{1}_{-2,0} \nonumber \\&\qquad + {\tiny \yng(1)} _{\,-1,-N}+{\tiny \yng(1)}_{\, 1,-N}+{\tiny \overline{ \yng(1)}}_{\, 1,N}+{\tiny \overline{ \yng(1)}}_{\, -1,N},
\end{align}
where the Coulomb branch is associated with the $U(1)_1$ subgroup and the components charged under $U(1)_1$ become massive along $\braket{Y_{SU(N-2)}^{bare}} \neq 0$.
Due to the mixed Chern-Simons term between $U(1)_1$ and $U(1)_2$, $Y_{SU(N-2)}^{bare}$ obtains a non-zero $U(1)_2$ charge (see Table \ref{SUSpin8electric}). In order to describe the moduli space in a gauge-invariant way, we need to define the dressed Coulomb branch operators
\begin{align}
Y^{dressed} &:=Y_{SU(N-2)}^{bare} ( {\tiny \overline{\yng(2)}}_{\,0,4})^{N-2}  \sim Y_{SU(N-2)}^{bare} \bar{S}^{N-2}  \\
Y_{W} &:=Y_{SU(N-2)}^{bare} ( {\tiny \overline{\yng(1)}}_{\,0,2})^{N-4}  {\tiny \overline{ \yng(1)}}_{\, 1,N}{\tiny \overline{ \yng(1)}}_{\, -1,N}  \sim Y_{SU(N-2)}^{bare}(\bar{S}Q)^{N-4} W_\alpha^2 \\
Y_{Q} &:=Y_{SU(N-2)}^{bare}  ( {\tiny \overline{\yng(2)}}_{\,0,4})^{N-2}   \mathbf{1}_{0,-2N+4} ( {\tiny \overline{\yng(1)}}_{\,0,2})^{N-2}     \sim Y_{SU(N-2)}^{bare}  \bar{S}^{N-2} \bar{S}(\bar{S}Q)^{N-2} 
\end{align}
Notice that these operators are not truncated at all since there is no superpotential in this electric theory. 
The quantum numbers of these operators are summarized in Table \ref{SUSpin8electric}.

\begin{table}[H]\caption{3d $\mathcal{N}=2$ $SU(N)$ gauge theory with $(N+4) \,  \protect\Young[0]{1}+\overline{ \protect\Young[0]{2}}$} 
\begin{center}
\scalebox{1}{
  \begin{tabular}{|c||c|c|c|c|c| } \hline
  &$SU(N)$&$SU(N+4)$&$U(1)$&$U(1)$& $U(1)_R$  \\ \hline
$Q$&${\tiny \yng(1)}$&${\tiny \yng(1)}$&1&0&$0$ \\   
 $\bar{S}$&${\tiny \overline{\yng(2)}}$&1&0&1&$0$ \\  \hline 
$M:=\bar{S}QQ$&1&${\tiny \yng(2)}$&2&1&0  \\
$B:=Q^N$&1&\scriptsize $\overline{\mbox{4-th anti-symm.}}$&$N$&0&0  \\
$U:= \det \bar{S}$&1&1&0&$N$&0  \\  \hline
\footnotesize $Y_{SU(N-2)}^{bare}$&\scriptsize $U(1)_2$: $-4(N-2)$&1&\footnotesize $-N-4$&$-N$&6 \\
\footnotesize $Y^{dressed}:=Y_{SU(N-2)}^{bare} \bar{S}^{N-2}$&1&1&\footnotesize $-N-4$&$-2$&6  \\ 
\footnotesize $Y_W:=Y_{SU(N-2)}^{bare}(\bar{S}Q)^{N-4} W_\alpha^2 $&1&\scriptsize $\overline{\mbox{8-th anti-symm.}}$&$-8$&$-4$&$8$\\
\footnotesize $Y_{Q}:=Y_{SU(N-2)}^{bare}  \bar{S}^{N-2} \bar{S}(\bar{S}Q)^{N-2}$&1&\scriptsize $\overline{\mbox{6-th anti-symm.}}$&$-6$&\footnotesize $N-3$&6  \\ \hline
  \end{tabular}}
  \end{center}\label{SUSpin8electric}
\end{table}

%%%%%%%%%%%%%%%%%%%%%%%%%%%%%%%%%%%%%%%%%%%%%%%%%%%%%%%%%%
The magnetic description is given by a 3d $\mathcal{N}=2$ $Spin(8)$ gauge theory with $F$ vectors $q$ and a spinor $s$ in addition to two gauge singlets $M$ and $Y^{dressed}$. The quantum numbers of the magnetic elementary fields are summarized in Table \ref{SUSpin8magnetic}. The theory has a tree-level superpotential
\begin{align}
W_{mag}=Mqq+Y^{dressed}ss, \label{Spin(8)swap}
\end{align}
which removes the magnetic mesons from the chiral ring elements. As for the Higgs branch, we have to include $sq^4s$ and $q^8$. The magnetic Coulomb branch $Y_{SO(6)}$ is completely the same as the previous analysis and corresponds to the breaking $so(8) \rightarrow so(6) \times u(1)$. Under the duality transformation, these gauge invariant operators are mapped as follows:
\begin{align}
B:=Q^N  \sim sq^4s,~~~Y_W \sim q^8,~~~U\sim Y_{SO(6)},~~~Y_Q \sim Y_{SO(6)}q^6,
\end{align}
which is consistent with our duality proposal (compare Table \ref{SUSpin8electric} and Table \ref{SUSpin8magnetic}).

\begin{table}[H]\caption{The magnetic description dual to Table \ref{SUSpin8electric}} 
\begin{center}
\scalebox{1}{
  \begin{tabular}{|c||c|c|c|c|c| } \hline
  &$Spin(8)$&$SU(N+4)$&$U(1)$&$U(1)$& $U(1)_R$  \\ \hline
$q$&$\mathbf{8}_v$&${\tiny \overline{\yng(1)}}$&$-1$&$-\frac{1}{2}$&$1$ \\   
 $s$&$\mathbf{8}_s$&1&$\frac{N+4}{2}$&1&$-2$ \\  
$M$&1&${\tiny \yng(2)}$&2&1&0  \\
$Y^{dressed}$&1&1&$-N-4$&$-2$&6  \\ \hline
$B \sim sq^4s$&1&\small $\overline{\mbox{4-th anti-symm.}}$&$N$&0&0  \\ 
$Y_W \sim q^8$ &1&\small $\overline{\mbox{8-th anti-symm.}}$&$-8$&$-4$&$8$  \\ \hline
$U\sim Y_{SO(6)}$&1&1&0&$N$&0 \\ 
$ Y_Q\sim Y_{SO(6)}q^6$&1&\small $\overline{\mbox{6-th anti-symm.}}$&$-6$&$N-3$&6  \\ \hline
  \end{tabular}}
  \end{center}\label{SUSpin8magnetic}
\end{table}

%%%%%%%%%%%%%%%%%%%%%%%%%%%%%%%%%%%%%%%%%%%%%%%%%%%%%%%%%%

Let us consider the $N=2$ case where the electric gauge group becomes $SU(2)$ and a symmetric representation is equivalent to adjoint. We should be careful that the Coulomb branch $Y_{SU(2)}$, which corresponds to the breaking $SU(2) \rightarrow U(1)$ (no $U(1)_2$), does not need ``dressing.'' Along the $Y_{SU(2)}$ direction, the adjoint matter reduces to $\mathbf{3} \rightarrow \mathbf{1}_{2} + \mathbf{1}_{0}+\mathbf{1}_{-2}$. Therefore, we can define the dressed operator $Y_{\bar{S}}:=Y_{SU(2)} \mathbf{1}_0 \sim Y_{SU(2)} \bar{S}$. The magnetic dual is given by the 3d $\mathcal{N}=2$ $Spin(8)$ gauge theory with six vectors and a spinor. The superpoetntial becomes $W_{mag}=Mqq +Y_{SU(2)}ss$. We can see the validity of this duality, for example, by flowing to an extended $\mathcal{N}=4$ supersymmetric theory. This can be achieved as follows: On the electric side, we add a cubic superpotential $\Delta W_{ele}= \sum_{i=1}^6 \bar{S}Q_iQ_i$ and the theory becomes a 3d $\mathcal{N}=4$ $SU(2)$ gauge theory with six doublets (three hypermultiplets in a doublet representation). The non-abelian global symmetry becomes $SO(6)$. On the magnetic side, the deformation corresponds to $\Delta W_{mag}=\sum_{i=1}^6 M_{ii}$ and breaks the gauge group to $Spin(2) \cong U(1)$. The low-energy superpotential becomes $W_{mag}= Y_{SU(2)} \sum_{i}^4  \bar{s}_{i} s^i$, which also has an extended $\mathcal{N}=4$ supersymmetry and an $SU(4)$ global symmetry. This duality was studied in \cite{Giacomelli:2019blm}. As a simple test of this 3d $\mathcal{N}=4$ duality\footnote{The electric and magnetic theories have been studied because of the fact that these theories have the same Higgs moduli space for an $SU(4) \cong SO(6)$ instanton \cite{Benvenuti:2010pq ,Hanany:2012dm}. Instead, by considering the $N$ $SU(4) \cong SO(6)$ instantons, we can conjecture the generalized 3d $\mathcal{N}=4$ duality as follows: The $SU(2)$ side is generalized to a 3d $\mathcal{N}=4$ $USp(2N)$ gauge theory with one (trace-full) antisymmetric and three fundamental hypermultiplets, which is dual to a 3d $\mathcal{N}=4$ $U(N)$ gauge theory with one adjoint and four fundamental hypermultiplets. By computing the superconformal indices, we checked the generalized duality for $N=1$ and $N=2$. We would like to thank Prof. Yuji Tachikawa for helpful comments on this generalization. See for example a recent related study \cite{Fan:2019jii}.}, we can easily see that these two theories have the same Hilbert series \cite{Pouliot:1998yv, Benvenuti:2010pq, Cremonesi:2013lqa, Fan:2019jii} for the Higgs and Coulomb branches:
\begin{align}
HS_{Higgs} &=\frac{1+9t^2 +9t^4+t^6}{(1-t^2)^6} \nonumber \\
&=1+15 t^2+84 t^4+300 t^6+825 t^8+1911 t^{10}+3920 t^{12}+7344 t^{14}+12825 t^{16}+\cdots  \\
HS_{Coulomb} &= \frac{1-t^4}{(1-t)(1-t^2)(1-t^2)}  \nonumber \\
&=1+t+3 t^2+3 t^3+5 t^4+5 t^5+7 t^6+7 t^7+9 t^8+9 t^9+11 t^{10}+11 t^{11}+13 t^{12}+\cdots
\end{align}
%

%%%%%%%%%%%%%%%%%%%%%%%%%%%%%%%%%%%%%%%%%%%%%%%%%%%%%%%%%%
\subsection{$Spin(8)$ with a spinor and a conjugate spinor \label{Spin8F11}}
%%%%%%%%%%%%%%%%%%%%%%%%%%%%%%%%%%%%%%%%%%%%%%%%%%%%%%%%%%
In this subsection, we propose a dual description for the 3d $\mathcal{N}=2$ $Spin(8)$ gauge theory with $F$ vectors $\mathbf{8_v}$, a spinor $\mathbf{8_s}$ and a conjugate spinor $\mathbf{8_c}$. It is known that the s-confinement phase appears when $F=4$ (see the reference \cite{Nii:2018wwj}). The corresponding 4d duality was studied in \cite{Cho:1997kr}. The Higgs branch operators are similar to the previous subsection and are defined as follows:
\begin{gather*}
M_{QQ}:=QQ,~~~M_{SS}:=SS,~~~M_{S'S'}:=SS,~~~P_1:=SQS' \\ 
P_3:=SQ^3S',~~~P_4:=SQ^4S,~~~P'_4:=S'Q^4S',~~~P_5:=SQ^5S',~~~P_7:=SQ^7S',~~~B:=Q^8.
\end{gather*}
On the second line, the color (and hence the flavor) indices of $Q$'s are anti-symmetrized. The mesonic operators on the first line (including $P_1$) will become elementary fields in the magnetic description discussed below.     

For the Coulomb branch of the moduli space, we need to take into account two types of flat directions due to two spinorial matters. This situation is different from the previous analysis. One of them is identical to the previous one $Y_{SO(6)}$ and corresponds to the breaking $so(8) \rightarrow so(6) \times u(1)$. Along the $Y_{SO(6)}$ flat direction, all the components of the spinor and conjugate spinor matters are massive while the vector remains massless as $\mathbf{6}_0$. Since there are massless vectors in the low-energy $SO(6)$ gauge theory, there is no runaway superpotential generated by the monopole-instantons of $U(1)^3 \subset SO(6)$. 
The additional Coulomb branch, which is denoted by $Z_{SO(4)}$, corresponds to the following breaking
\begin{align}
so(8) & \rightarrow so(4) \times su(2) \times u(1) \\
\mathbf{8_v} & \rightarrow (\mathbf{4}, \mathbf{1} )_0 + (\mathbf{1}, \mathbf{2} )_{\pm 1} \\
\mathbf{8_s} & \rightarrow (\mathbf{2}, \mathbf{2} )_0 + (\mathbf{2'}, \mathbf{1} )_{\pm 1} \\
\mathbf{8_c} & \rightarrow (\mathbf{2'}, \mathbf{2} )_0 + (\mathbf{2}, \mathbf{1} )_{\pm 1}
\end{align}
where the chiral superfield $Z_{SO(4)}$ can be obtained by dualizing the unbroken $U(1)$ vector superfield. In $so(4) \cong su(2) \times su(2)$, there are two spinor representations, $\mathbf{2}$ and $\mathbf{2'}$.
The vacuum of the low-energy $SO(4)$ dynamics can be made supersymmetric by the massless components $(\mathbf{4}, \mathbf{1} )_0$ while the vacuum of the $SU(2)$ gauge theory can be stable due to $(\mathbf{2}, \mathbf{2} )_0$ and $(\mathbf{2'}, \mathbf{2} )_0$. For theories with a single spinor (or a conjugate spinor), the low-energy $SU(2)$ dynamics has a quantum-deformed constraint between the Coulomb and Higgs branch coordinates and then the Coulomb branch can be eliminated. Therefore, the theories with more than one (conjugate) spinor has this flat direction $Z_{SO(4)}$. By using the bare monopole operators $Y_{SO(6)}$ and $Z_{SO(4)}$, we can also define the baryon-monopole operators
\begin{align}
Y_{SO(6)}^Q&:=  Y_{SO(6)}  \left( \mathbf{6}_0 \right)^6  \nonumber \\
& \sim Y_{SO(6)}  Q^6 \\
Z_{SO(4)}^Q&:=Z_{SO(4)} \left((\mathbf{4}, \mathbf{1} )_0  \right)^4 \nonumber \\
&\sim Z_{SO(4)}  Q^4,
\end{align}
where the color indices of $Q^6$ and $Q^4$ are contracted by epsilon tensors of the $SO(6)$ and $SO(4)$ groups, respectively. The quantum numbers of these operators are summarized in Table \ref{Spin8F11electric}.

\if0
Although we don't study the superconformal indices for $Spin(8)$ cases in this paper, we briefly explain the SCI point of view of the Coulomb branch. The superconformal index is a summation over the GNO charges $(m_1,m_2,m_3,m_4)$. $Y_{SO(6)}$ corresponds to $(1,0,0,0)$ and $Z_{SO(4)}$ is $(1,1,0,0)$. 
\fi

\begin{table}[H]\caption{3d $\mathcal{N}=2$ $Spin(8)$ gauge theory with $(N_v, N_s, N_c)=(F,1,1)$} 
\begin{center}
\scalebox{1}{
  \begin{tabular}{|c||c|c|c|c|c|c| } \hline
  &$Spin(8)$&$SU(F)$&$U(1)$&$U(1)$&$U(1)$& $U(1)_R$  \\ \hline
$Q$&$\mathbf{8_v}$&${\tiny \yng(1)}$&1&0&0&0 \\
$S$&$\mathbf{8_s}$&1&0&1&0&0  \\
$S'$&$\mathbf{8_c}$&1&0&0&1&0  \\ \hline
$M_{QQ}:=QQ$&1&${\tiny \yng(2)}$&2&0&0&0  \\
$M_{SS}:=SS$&1&1&0&2&0&0  \\
$M_{S'S'}:=S'S'$&1&1&0&0&2&0  \\
$P_1:=SQS'$&1&${\tiny \yng(1)}$&1&1&1&0  \\
$P_3:=SQ^3S'$&1&\scriptsize 3rd anti-symm.&3&1&1&0  \\
$P_4:=SQ^4S$&1&\scriptsize 4-th anti-symm.&4&2&0&0  \\
$P'_4:=S'Q^4S'$&1&\scriptsize 4-th anti-symm.&4&0&2&0  \\ 
$P_5:=SQ^5S'$&1&\scriptsize 5-th anti-symm.&5&1&1&0  \\
$P_{7}:=SQ^7S'$&1&\scriptsize 7-th anti-symm.&7&1&1&0 \\
$B:=Q^8$&1&\scriptsize 8-th anti-symm.&8&0&0&0  \\ \hline
$Y_{SO(6)}$&1&1&$-2F$&$-4$&$-4$&$2F-4$  \\
$Z_{SO(4)}$&1&1&$-2F$&$-2$&$-2$&$2F-6$\\
$Y_{SO(6)}^Q:= Y_{SO(6)} Q^6$&1&\scriptsize 6-th anti-symm.&$6-2F$&$-4$&$-4$&$2F-4$  \\
$Z_{SO(4)}^Q:=Z_{SO(4)} Q^4$&1&\scriptsize 4-th anti-symm.&$4-2F$&$-2$&$-2$&$2F-6$  \\ \hline
  \end{tabular}}
  \end{center}\label{Spin8F11electric}
\end{table}

%%%%%%%%%%%%%%%%%%%%%%%%%%%%%%%%%%%%%%%%%%%%%%%%%%%%%%%%%%
The magnetic description is given by a 3d $\mathcal{N}=2$ $SU(F-3)$ gauge theory with $F+2$ fundamental matters, an anti-fundamental matter and a symmetric-bar tensor. In addition to these quarks, the theory contains four gauge-singlets $M_{QQ}, M_{SS}, M_{S'S'}$ and $P_1$ which are identified with the electric meson operators in Table \ref{Spin8F11electric}. The theory has a tree-level superpotential 
\begin{align}
W_{mag} = M_{QQ}\bar{s}qq+\bar{s}q'q''+M_{SS} q''\bar{q} +M_{S'S'} q' \bar{q} +P_1q \bar{q}+\tilde{Y}^{dressed}, \label{WmagSpin8F11}
\end{align}
which decomposes the fundamental matters into three groups $q, q'$ and $q''$. The global symmetry reduces to $SU(F) \times U(1) \times U(1) \times U(1) \times U(1)_R$ which is identical to the electric symmetry. The charge assignment of the dual fields is completely fixed by the above superpotential as in Table \ref{Spin8F11magnetic}. Therefore, the matching of the gauge-invariant operators becomes a non-trivial test of the duality. We begin with the study of mapping the Higgs branch operators. On the magnetic side, we can define the following gauge-invariant coordinates and find the matching
\begin{gather*}
P_3 \sim q^{F-3},~~~P_4 \sim q^{F-4} q',~~~P_4 \sim q^{F-4} q',~~~P'_4 \sim q^{F-4} q'' \\
P_5 \sim q^{F-5}q'q'',~~~Z_{SO(4)} \sim \bar{s}^{F-3},~~~Y_{SO(6)} \sim \bar{s}^{F-4} \bar{q}^2,~~~Z_{SO(4)}^Q \sim (\bar{s}q)^{F-4} \bar{q},
\end{gather*}
where the operators including $\bar{s}$ are transformed into the electric Coulomb branches under the duality. Notice that $(\bar{s}q)^{F-3} \sim \det \bar{s} \, q^{F-3}$ cannot be an independent operator and also that $\bar{s}q'$ and $\bar{s}q''$ are eliminated from the chiral ring elements due to the $F$-flatness conditions of the superpotential. All the magnetic mesons are lifted due to the $F$-flatness conditions as well. 

We will next consider the Coulomb branch: When the magnetic Coulomb branch $\tilde{Y}_{SU(F-5)}^{bare}$ obtains a non-zero expectation value, the gauge group is spontaneously broken to
\begin{align}
SU(F-3) & \rightarrow SU(F-5) \times U(1)_1 \times U(1)_2 \\ 
{\tiny \yng(1)}  & \rightarrow  {\tiny \yng(1)}_{\, 0,-2} +\mathbf{1}_{1,F-5}+\mathbf{1}_{-1,F-5}\\
{\tiny \overline{\yng(1)}}  & \rightarrow  {\tiny \overline{\yng(1)}}_{\, 0,2} +\mathbf{1}_{-1,-(F-5)}+\mathbf{1}_{1,-(F-5)} \\
{\tiny \overline{ \yng(2)}} & \rightarrow {\tiny \overline{ \yng(2)}}_{\, 0,4} +{\tiny \overline{ \yng(1)}}_{\,-1,-F+7}  +{\tiny \overline{ \yng(1)}}_{\, 1,-F+7} \nonumber \\&\qquad \qquad  +\mathbf{1}_{-2,-2F+10} +\mathbf{1}_{2,-2F+10}+\mathbf{1}_{0,-2F+10}   \\
\mathbf{adj.} & \rightarrow  \mathbf{adj.}_{0,0}+\mathbf{1}_{0,0}+\mathbf{1}_{0,0}+   \mathbf{1}_{2,0}+\mathbf{1}_{-2,0} \nonumber \\&\qquad \qquad  + {\tiny \yng(1)} _{\,-1,-F+3}+{\tiny \yng(1)}_{\, 1,-F+3}+{\tiny \overline{ \yng(1)}}_{\, 1,F-3}+{\tiny \overline{ \yng(1)}}_{\, -1,F-3},
\end{align}
where $\tilde{Y}_{SU(F-5)}^{bare}$ corresponds to the $U(1)_1$ generator. The decomposed components obtain real masses proportional to their $U(1)_1$ charges. By integrating out the massive components, this branch obtains a non-zero mixed Chern-Simons term $k_{eff}^{U(1)_1 U(1)_2}=4(F-5)$. As a result, the bare Coulomb branch $\tilde{Y}_{SU(F-5)}^{bare}$ has a $U(1)_2$ charge $-k_{eff}^{U(1)_1 U(1)_2}$. See Table \ref{Spin8F11magnetic} for quantum numbers of $\tilde{Y}_{SU(F-5)}^{bare}$.

In order to describe the Coulomb branch, we need gauge-invariant operators consisting of the Coulomb branch. By combining the bare Coulomb and Higgs branch operators, we can define the following (gauge-invariant) baryon-monopole operators
\begin{align}
\tilde{Y}^{dressed}  &:=\tilde{Y}_{SU(F-5)}^{bare} ({\tiny \overline{ \yng(2)}}_{\, 0,4})^{F-5} \nonumber \\
&\sim   \tilde{Y}_{SU(F-5)}^{bare} \bar{s}^{F-5} \\
Y_{SO(6)}^Q & :=  \tilde{Y}_{SU(F-5)}^{bare}  ({\tiny \overline{ \yng(2)}}_{\, 0,4})^{F-6}  ({\tiny \overline{ \yng(1)}}_{\, 0,2})^{2} \mathbf{1}_{0,-2F+10} ({\tiny \overline{ \yng(1)}}_{\, 0,2})^{F-6} {\tiny \overline{ \yng(1)}}_{\, 0,2}   \nonumber \\
& \sim \tilde{Y}_{SU(F-5)}^{bare}  \bar{s}^{F-6} \bar{q}^2 \bar{s}(\bar{s}q)^{F-6}\bar{q} \\
P_7&:=\tilde{Y}_{SU(F-5)}^{bare}({\tiny \overline{ \yng(1)}}_{\, 0,2})^{F-7} {\tiny \overline{ \yng(1)}}_{\, 1,F-3}{\tiny \overline{ \yng(1)}}_{\, -1,F-3} \nonumber \\
&  \sim \tilde{Y}_{SU(F-5)}^{bare} (\bar{s}q)^{F-7} \tilde{W}_\alpha^2  \\
B&:=  \tilde{Y}_{SU(F-5)}^{bare}   ({\tiny \overline{ \yng(1)}}_{\, 0,2})^{F-8} {\tiny \overline{ \yng(1)}}_{\, 0,2} {\tiny \overline{ \yng(1)}}_{\, 1,F-3}{\tiny \overline{ \yng(1)}}_{\, -1,F-3} \nonumber \\
& \sim \tilde{Y}_{SU(F-5)}^{bare} (\bar{s}q)^{F-8} \bar{q} \tilde{W}_\alpha^2 
\end{align}
where the color indices of the mattter chiral superfields are contracted by epsilon tensors of the unbroken $SU(F-5)$ group. $\tilde{Y}^{dressed}$ is lifted due to the superpotential. The quantum numbers of these operators are summarized in Tbale \ref{Spin8F11magnetic} and this confirms the operator mapping. This can be seen as a validity check of our duality proposal.

\begin{table}[H]\caption{The magnetic $SU(F-3)$ gauge theory dual to Table \ref{Spin8F11electric}} 
\begin{center}
\scalebox{0.83}{
  \begin{tabular}{|c||c|c|c|c|c|c| } \hline
  &$SU(F-3)$&$SU(F)$&$U(1)$&$U(1)$&$U(1)$& $U(1)_R$  \\ \hline
$q$&${\tiny \yng(1)}$&${\tiny \overline{\yng(1)}}$&$\frac{3}{F-3}$&$\frac{1}{F-3}$&$\frac{1}{F-3}$&0  \\
$q'$&${\tiny \yng(1)}$&1&$\frac{F}{F-3}$&\footnotesize $1+\frac{1}{F-3}$&\footnotesize $-1+\frac{1}{F-3}$&0  \\
$q''$&${\tiny \yng(1)}$&1&$\frac{F}{F-3}$&\footnotesize $-1+\frac{1}{F-3}$&\footnotesize $1+\frac{1}{F-3}$&0  \\
$\bar{q}$&${\tiny \overline{\yng(1)}}$&1&$-\frac{F}{F-3}$&\footnotesize $-1-\frac{1}{F-3}$&\footnotesize$-1-\frac{1}{F-3}$&2  \\ 
$\bar{s}$&${\tiny \overline{\yng(2)}}$&1&$-\frac{2F}{F-3}$&$-\frac{2}{F-3}$&$-\frac{2}{F-3}$&2  \\
$M_{QQ}$&1&${\tiny \yng(2)}$&2&0&0&0  \\
$M_{SS}$&1&1&0&2&0&0  \\
$M_{S'S'}$&1&1&0&0&2&0  \\
$P_1$&1&${\tiny \yng(1)}$&1&1&1&0  \\  \hline
$P_3 \sim q^{F-3}$&1&\scriptsize 3-th anti-symm.&3&1&1&0  \\
$P_4 \sim q^{F-4} q'$&1&\scriptsize 4-th anti-symm.&4&2&0&0  \\
$P'_4 \sim q^{F-4} q''$&1&\scriptsize 4-th anti-symm.&4&0&2&0  \\ 
$P_5 \sim q^{F-5}q'q''$&1&\scriptsize 5-th anti-symm.&5&1&1&0  \\ 
$Z_{SO(4)} \sim \bar{s}^{F-3}$&1&1&$-2F$&$-2$&$-2$&\footnotesize $2F-6$ \\
$Y_{SO(6)} \sim \bar{s}^{F-4} \bar{q}^2$&1&1&$-2F$&$-4$&$-4$&\footnotesize $2F-4$  \\
$Z_{SO(4)}^Q \sim (\bar{s}q)^{F-4} \bar{q}$&1&\scriptsize 4-th anti-symm.&$4-2F$&$-2$&$-2$&\footnotesize $2F-6$  \\ \hline
$\tilde{Y}_{SU(F-5)}^{bare}$&\scriptsize $U(1)_2$: $-4(F-5)$&1& \scriptsize$2F-1-\frac{3F+3}{F-3}$& \scriptsize $3-\frac{F+1}{F-3}$&\scriptsize $3-\frac{F+1}{F-3}$&\scriptsize $-2F+12$  \\
\footnotesize $\tilde{Y}^{dressed}:= \tilde{Y}_{SU(F-5)}^{bare} \bar{s}^{F-5}$&1&1&0&0&0&2  \\
\footnotesize $Y_{SO(6)}^Q \sim  \tilde{Y}_{SU(F-5)}^{bare}  \bar{s}^{F-6} \bar{q}^2 \bar{s}(\bar{s}q)^{F-6}\bar{q} $&1&\scriptsize 6-th anti-symm.&\footnotesize $6-2F$&$-4$&$-4$&\footnotesize $2F-4$  \\ 
\footnotesize $P_7\sim \tilde{Y}_{SU(F-5)}^{bare} (\bar{s}q)^{F-7} \tilde{W}_\alpha^2 $&1&\scriptsize 7-th anti-symm.&7&1&1&0  \\
\footnotesize $B \sim \tilde{Y}_{SU(F-5)}^{bare} (\bar{s}q)^{F-8} \bar{q} \tilde{W}_\alpha^2 $&1&\scriptsize 8-th anti-symm.&8&0&0&0   \\ \hline
%&&&&&&  \\ \hline
  \end{tabular}}
  \end{center}\label{Spin8F11magnetic}
\end{table}

%%%%%%%%%%%%%%%%%%%%%%%%%%%%%%%%%%%%%%%%%%%%%%%%%%%%%%%%%%
Several consistency checks are in order: First, this duality can go back to the $Spin(8)$ duality with a spinor matter by introducing a complex mass $W= m S'S' $ to the conjugate spinor and integrating out it. On the magnetic side, the same mass leads to $\braket{q'\bar{q}}=-m$ which breaks the gauge group to $SU(F-4)$ and one flavor ($q'$ and $\bar{q}$) is eaten via the Higgs mechanism. By solving the $F$-flatness conditions, we find that $q^{F-3}, q''$ and $\bar{s}_{F-3,a} \, (a=1,\cdots,F-4)$ are massive and that the low-energy theory has the following massless degrees of freedom
\begin{align}
\hat{\bar{s}} &:= \bar{s}_{ab}~~(a,b=1.\cdots,F-4) \\
\bar{s}_{singlet} &:= \bar{s}_{F-3,F-3}=-M_{SS} \\
\hat{q}_i^a &:=q_i^{a}~~(a=1,\cdots,F-4.~~i=1,\cdots,F),
\end{align}
where $i$ denotes the flavor index and $a,b$ are color indices. The component $\bar{s}_{F-3,F-3}$ is identified with the spinor meson $M_{SS}$. Therefore, the high-energy dressed operator $\tilde{Y}^{dressed}:=\tilde{Y}_{SU(F-5)}^{bare} \bar{s}^{F-5} $ flows to $M_{SS} \tilde{Y}^{dressed}_{low}$, where $\tilde{Y}^{dressed}_{low}:= \tilde{Y}_{SU(F-3)}^{bare} \hat{\bar{s}}^{F-6} $ .
 The low-energy superpotential becomes
\begin{align}
W_{mag}^{low} = M_{QQ} \hat{\bar{s}} \hat{q} \hat{q} + M_{SS} \tilde{Y}^{dressed}_{low},
\end{align}
which reproduces the $Spin(8)$ duality with $F$ vectors and a spinor.

Next, let us consider adding a non-zero vev to $M_{QQ}$ such that $\mathrm{rank} \braket{M_{QQ}}=1$. On the electric side, the gauge group is broken to $Spin(7)$. The spinor $\mathbf{8_s}$ (and also $\mathbf{8_c}$) reduces to a $Spin(7)$ spinor. We thus obtain a 3d $\mathcal{N}=2$ $Spin(7)$ gauge theory with $F$ vectors and two spinors, where we shifted $F \rightarrow F+1$ for simplicity. On the magnetic side, $\mathrm{rank} \braket{M_{QQ}}=1$ decomposes $q$ into $F-1$ quarks and a single one $q'''$. Since there is no superpotential interaction distinguishing $q', q''$ from $q'''$, we have an enhanced $SO(3) \cong SU(2)$ symmetry rotating the two spinors in the electric theory.  In this way, we precisely obtain the $Spin(7)$ duality with two spinors, which was studied in the previous section.

Similarly, we can study the flat direction spanned by $\braket{M_{S'S'}}$ which spontaneously breaks $Spin(8)$ into $Spin(7)$. In a low-energy limit of the electric side, we obtain a $Spin(7)$ gauge theory with $F+1$ spinors. Notice that $\mathbf{8_v}$ and $\mathbf{8_s}$ reduce to a $Spin(7)$ spinor representation. On the magnetic side, $q'$ and $\bar{q}$ become massive and integrated out. The low-energy limit becomes an $SU(F-3)$ gauge theory with $F+1$ fundamental matters and a symmetric-bar tensor. The low-energy superpotential becomes
\begin{align}
W  &= M_{QQ} \bar{s}qq +M_{SS} \bar{s}q'' q'' +P_1 \bar{s} qq''+ \tilde{Y}^{dressed} \\
&=\hat{M}_{QQ} \bar{s} \hat{q} \hat{q} +\tilde{Y}^{dressed},
\end{align}
where we used the equation of motion of $q'$ to eliminate $q'$ and $\bar{q}$. On the second line, we combined the $F+1$ fundamental matters into a single field $\hat{q}$. $M_{QQ},M_{SS}$ and $P_1$ are also combined into $\hat{M}_{QQ}$. This reproduces the $Spin(7)$ duality studied in \cite{Nii:2019wjz}. These deformations confirm the validity of our duality proposal.

%%%%%%%%%%%%%%%%%%%%%%%%%%%%%%%%%%%%%%%%%%%%%%%%%%%%%%%%%%
%%%%%%%%%%%%%%%%%%%%%%%%%%%%%%%%%%%%%%%%%%%%%%%%%%%%%%%%%%
\section{3d $Spin(9)$ Seiberg duality}
%%%%%%%%%%%%%%%%%%%%%%%%%%%%%%%%%%%%%%%%%%%%%%%%%%%%%%%%%%
%%%%%%%%%%%%%%%%%%%%%%%%%%%%%%%%%%%%%%%%%%%%%%%%%%%%%%%%%%
In this section, we present the $Spin(9)$ Seiberg duality with vector and spinor matters. The corresponding 4d duality was proposed in \cite{Cho:1997kr}. The electric description is given by a 3d $\mathcal{N}=2$ $Spin(9)$ gauge theory with $F$ vector matters and a single spinor matter. The s-confinement phase appears when $F=5$ \cite{Nii:2018wwj}. We here generalize the analysis in \cite{Nii:2018wwj} for large flavor cases $F \ge 5$. The Higgs branch is described by mesonic and baryonic operators
\begin{gather*}
M_{QQ}:=QQ,~~~M_{SS}:=SS,~~~P_1:= SQS \\
P_4:=SQ^4S,~~~P_5:=SQ^5S,~~~P_8:=SQ^8S,~~~B:=Q^9,
\end{gather*}
where $Q$'s on the second line are anti-symmetrized by anti-symmetric combinations of $Spin(9)$ gamma matrices. In a $Spin(9)$ group, a mass operator for a single spinor is available. The operators on the first line will become elementary fields in a magnetic theory. There are two types of Coulomb branches that we have to take into account \cite{Aharony:2013kma, Nii:2018wwj}. The first (bare) Coulomb branch, denoted by $Y_{SO(7)}$, induces the following symmetry breaking
\begin{align}
so(9) & \rightarrow so(7) \times u(1) \\
\mathbf{9} & \rightarrow  \mathbf{7}_0+\mathbf{1}_2+\mathbf{1}_{-2} \\
\mathbf{16} & \rightarrow \mathbf{8}_1 +\mathbf{8}_{-1}
\end{align}
This flat direction is quantum-mechanically stable and supersymmetric since there are massless components $\mathbf{7}_0$ from the vector matters which can forbid the monopoles from generating instabilities of runaway potential. The low-energy $Spin(7)$ gauge theory with more than five vectors has supersymmetric vacua whose Coulomb branch is not removed \cite{Aharony:2011ci}. We thus need to introduce $Y_{SO(7)}$ for parametrizing the flat direction of the $Spin(9)$ gauge theory with $F \ge 6$ vectors. One can also define a baryon-monopole operator \cite{Aharony:2013kma}
\begin{align}
Y_{SO(7)}^Q := Y_{SO(7)} ( \mathbf{7}_0)^7  \sim Y_{SO(7)} Q^7,
\end{align}
where the color indices of $Q^7$ are contracted by a $Spin(7)$ epsilon tensor. This operator is available for $F \ge 7$.

There is another Coulomb branch possible for $Spin(9)$ theories with spinor matters. When the second Coulomb branch, denoted by $Z_{SO(5)}$, obtains a non-zero expectation value, the gauge group is spontaneously broken to
\begin{align}
so(9) & \rightarrow so(5) \times su(2) \times u(1) \\
\mathbf{9} & \rightarrow  (\mathbf{5},\mathbf{1})_0+  (\mathbf{1},\mathbf{2})_{\pm 1}\\
\mathbf{16} & \rightarrow  (\mathbf{4},\mathbf{2})_0+ (\mathbf{4},\mathbf{1})_{\pm 1},
\end{align}
where $Z_{SO(5)}$ is associated with the unbroken $U(1)$ subgroup.  
Since there are two non-abelian factors, these two gauge dynamics must have supersymmetric vacua. Otherwise, this classical flat direction will be lifted due to a non-perturbative dynamics of $SO(5)$ or $SU(2)$.
Along the $Z_{SO(5)}$ Coulomb branch, the vector representation reduces to a massless vector of $SO(5)$ while the spinor becomes $(\mathbf{4},\mathbf{2})_0$. All the other components are massive and integrated out. This flat direction can be quantum-mechanically stable since there are massless dynamical quarks for both of the unbroken gauge groups. For a $Spin(9)$ gauge theory only with vector matters, the $SU(2)$ subgroup is unstable due to the monopole (runaway) superpotential \cite{Aharony:1997bx, deBoer:1997kr} from a compact $U(1) \subset SU(2)$. As in the case of $Y_{SO(7)}$, one can define the baryon-monopole operator
\begin{align}
Z_{SO(5)}^Q :=Z_{SO(5)}  \left(  (\mathbf{5},\mathbf{1})_0 \right)^5 \sim Z_{SO(5)} Q^5,
\end{align}
where the color indices of $Q$ are contracted by an epsilon tensor of $SO(5)$. This operator is available for $F \ge 5$. These dressed operators can be regarded as 3d counterparts of the exotic baryons in the 4d $Spin(N)$ gauge theory \cite{Aharony:2013kma}.

%%%%%%%%%%%%%%%%%%%%%%%%%%%%%%%%%%%%%%%%%%%%%%%%%%%%%%%%%%
Based on the analysis of the Coulomb branch above, we can write down the effective superpotential for small $F$ cases. For $F=5$, we need to introduce the dressed monopole $Z_{SO(5)}^Q$ while only $Z_{SO(5)}$ can appear for $F \le 4$. The global symmetries in Table \ref{Spin9electric} determine the effective potentials as follows: 
\begin{gather*}
W_{F=5}^{eff} = Z_{SO(5)} (M_{QQ}^5 M_{SS}^2 +M_{QQ}^4 P_1^2 +M_{QQ}P_4^2 +P_5^2) +Z_{SO(5)}^{Q} (M_{SS} P_5 +P_1P_4) \\
W_{F=4}^{eff} = \lambda \left[Z_{SO(5)} (M_{QQ}^4 M_{SS}^2 +M_{QQ}^3 P_1^2+P_4^2) -1   \right]\\
W_{F=3}^{eff} = \frac{1}{Z_{SO(5)} (M_{QQ}^3 M_{SS}^2 +M_{QQ}^2 P_1^2 )},~~~W_{F=2}^{eff} = \left( \frac{1}{Z_{SO(5)} (M_{QQ}^2 M_{SS}^2 +M_{QQ} P_1^2)} \right)^{\frac{1}{2}} \\
W_{F=1}^{eff} = \left( \frac{1}{Z_{SO(5)} (M_{QQ} M_{SS}^2+P_1^2)} \right)^{\frac{1}{3}},~~~ W_{F=0}^{eff} = \left( \frac{1}{Z_{SO(5)} M_{SS}^2} \right)^{\frac{1}{2}}
\end{gather*}
where $\lambda$ is a Lagrange multiplier. 
For $F \le 3$, the effective superpotential becomes runaway and there is no stable supersymmetric vacuum. For $F=4$, there is one quantum constraint between the Higgs and Coulomb branch coordinates, where the semi-classical region of the Higgs branch is connected with the strongly-coupled region of the Coulomb branch and vice versa. For $F=5$, the theory exhibits s-confinement as studied in \cite{Nii:2018wwj}.

\begin{table}[H]\caption{3d $\mathcal{N}=2$ $Spin(9)$ gauge theory with $F$ vectors and a spinor} 
\begin{center}
\scalebox{1}{
  \begin{tabular}{|c||c|c|c|c|c| } \hline
  &$Spin(9)$&$SU(F)$&$U(1)$&$U(1)$& $U(1)_R$  \\ \hline
$Q$&$\mathbf{9}$&${\tiny \yng(1)}$&1&0&$0$ \\   
 $S$&$\mathbf{16}$&1&0&1&$0$ \\  \hline 
$M_{QQ}:=Q^2$ &1&${\tiny \yng(2)}$&2&0&0  \\
$M_{SS}:=S^2$ &1&1&0&2&0  \\
$P_1:= SQS$&1&${\tiny \yng(1)}$&1&2&0 \\
$P_4:=SQ^4S$&1&\footnotesize 4-th anti-symm.&4&2&0  \\
$P_5:=SQ^5S$&1&\footnotesize 5-th anti-symm.&5&2&0  \\ 
$P_8:=SQ^8S$&1&\footnotesize 8-th anti-symm.&8&2&0  \\
$B:=Q^9$&1&\footnotesize 9-th anti-symm.&9&0&0  \\ \hline
%$B:=Q^7$&1&7-th anti-symm. &7&0&0  \\
%$P_3:=SQ^3S$ &1&${\tiny \yng(1,1,1)}$&3&2&0  \\[5pt]
$Y_{SO(7)}$&1&1&$-2F$&$-8$&$2F-6$  \\
$Z_{SO(5)}$&1&1&$-2F$&$-4$&$2F-8$  \\
$Y_{SO(7)}^Q:=Y_{SO(7)}Q^7$&1&\footnotesize 7-th anti-symm.&$7-2F$&$-8$&$2F-6$  \\
$Z_{SO(5)}^{Q}:=Z_{SO(5)}Q^5$&1&\footnotesize 5-th anti-symm.&$5-2F$&$-4$&$2F-8$ \\ \hline
  \end{tabular}}
  \end{center}\label{Spin9electric}
\end{table}

%%%%%%%%%%%%%%%%%%%%%%%%%%%%%%%%%%%%%%%%%%%%%%%%%%%%%%%%%%
The magnetic description is a 3d $\mathcal{N}=2$ $SU(F-4)$ gauge theory with $F+1$ fundamental matters, an anti-fundamental matter and a symmetric-bar tensor. As advertised above, there are also three gauge-singlet chiral superfields $M_{QQ}, M_{SS}$ and $P_1$, which are identified with the electric counterparts appearing in Table \ref{Spin9electric}. The theory includes a tree-level superpotential  
\begin{align}
W_{mag}=P_1 q \bar{q} +M_{SS} q' \bar{q} +M_{QQ}\bar{s}qq +\bar{s}q'q' +\tilde{Y}^{dressed},
\end{align}
which decomposes the fundamental matters into $F$ quarks $q$ and a single quark $q'$. Thus, the global symmetry is reduced to $SU(F) \times U(1) \times U(1) \times U(1)_R$. The charge assignment of the magnetic elementary fields is fixed by the superpotential. The Higgs branch operators in the magnetic theory are defined as
\begin{gather*}
Z_{SO(5)} \sim \det \bar{s},~~~Y_{SO(7)} \sim \bar{s}^{F-5} \bar{q}^2,~~~Z_{SO(5),Q} \sim (\bar{s}q)^{F-5} \bar{q} \\
P_4 \sim q^{F-4},~~~P_5 \sim q^{F-5} q',
\end{gather*}
where we also indicated the operator matching under the duality. Notice that the determinant $\det \bar{s}$ is not truncated by the superpotential as opposed to the 4d case. Notice also that $(\bar{s} q)^{F-4}$ cannot be an independent operator. The (dressed) Coulomb branch operators along $Z_{SO(5)}$ are mapped to the Higgs branch operators under the duality transformation. Since the charge assignment is fixed by the superpotential, the operator matching above can be regarded as a non-trivial test of the duality.  

We can also find the matching of the remaining operators: The magnetic Coulomb branch, denoted by $\tilde{Y}_{SU(F-6)}^{bare}$, corresponds to the gauge symmetry breaking
\begin{align}
SU(F-4) & \rightarrow SU(F-6) \times U(1)_1 \times U(1)_2 \\
{\tiny \yng(1)}  & \rightarrow  {\tiny \yng(1)}_{\, 0,-2} +\mathbf{1}_{1,F-6}+\mathbf{1}_{-1,F-6}\\
{\tiny \overline{\yng(1)}}  & \rightarrow  {\tiny \overline{\yng(1)}}_{\, 0,2} +\mathbf{1}_{-1,-F+6}+\mathbf{1}_{1,-F+6}\\
{\tiny \overline{ \yng(2)}} & \rightarrow {\tiny \overline{ \yng(2)}}_{\, 0,4} +{\tiny \overline{ \yng(1)}}_{\,-1,-F+8}  +{\tiny \overline{ \yng(1)}}_{\, 1,-F+8} +\mathbf{1}_{-2,-2F+12} +\mathbf{1}_{2,-2F+12}+\mathbf{1}_{0,-2F+12}   \\
\mathbf{adj.} & \rightarrow  \mathbf{adj.}_{0,0}+\mathbf{1}_{0,0}+\mathbf{1}_{0,0}+   \mathbf{1}_{2,0}+\mathbf{1}_{-2,0} \nonumber \\&\qquad + {\tiny \yng(1)} _{\,-1,-F+4}+{\tiny \yng(1)}_{\, 1,-F+4}+{\tiny \overline{ \yng(1)}}_{\, 1,F-4}+{\tiny \overline{ \yng(1)}}_{\, -1,F-4},
\end{align}
where the Coulomb branch corresponds to the unbroken $U(1)_1$ generator.
The components charged under the $U(1)_1$ subgroup are massive and integrated out, which generates a non-zero mixed Chern-Simons term $k_{eff}^{U(1)_1 U(1)_2}=4(F-6)$. As a result, the bare Coulomb branch obtains a $U(1)_2$ charge $-k_{eff}^{U(1)_1 U(1)_2}$ \cite{Intriligator:2013lca}. In order to construct gauge-invariant operators along the Coulomb branch, we need to define baryon-monopole operators \cite{Aharony:2015pla, Csaki:2014cwa, Nii:2018bgf}
\begin{align}
\tilde{Y}^{dressed} &:=\tilde{Y}_{SU(F-6)}^{bare} ( {\tiny \overline{ \yng(2)}}_{\, 0,4})^{F-6}   \sim  \tilde{Y}_{SU(F-6)}^{bare} \bar{s}^{F-6}  \\
Y_{SO(7)}^Q &:=    \tilde{Y}^{bare}_{SU(F-6)} ({\tiny \overline{ \yng(2)}}_{\, 0,4})^{F-7} ({\tiny \overline{ \yng(1)}}_{\, 0,2})^2 \mathbf{1}_{0,-2F+12}  ({\tiny \overline{ \yng(1)}}_{\, 0,2})^{F-7} {\tiny \overline{ \yng(1)}}_{\, 0,2}  \\
& \sim \tilde{Y}^{bare}_{SU(F-6)} \bar{s}^{F-7} \bar{q}^2 \bar{s}(\bar{s}q)^{F-7} \bar{q}  \\
B  &:=  \tilde{Y}_{SU(F-6)}^{bare} ({\tiny \overline{ \yng(1)}}_{\, 0,2} )^{F-9}{\tiny \overline{ \yng(1)}}_{\, 0,2}  {\tiny \overline{ \yng(1)}}_{\, 1,F-4}{\tiny \overline{ \yng(1)}}_{\, -1,F-4}  \sim  \tilde{Y}_{SU(F-6)}^{bare} (\bar{s}q)^{F-9} \bar{q} \tilde{W}_\alpha^2,
\end{align}
where the color indices are all contracted by epsilon tensors of the unbroken $SU(F-6)$ gauge group. 
The global charges of these dressed operators are computed as in Table \ref{Spin9magnetic} and are consistent with the electric counterparts. The dressed operator $\tilde{Y}^{dressed}$ is truncated due to the superpotential.

\begin{table}[H]\caption{The magnetic $SU(F-4)$ gauge theory dual to Table \ref{Spin9electric}} 
\begin{center}
\scalebox{0.9}{
  \begin{tabular}{|c||c|c|c|c|c| } \hline
  &$SU(F-4)$&$SU(F)$&$U(1)$&$U(1)$& $U(1)_R$  \\ \hline
$q$&${\tiny \yng(1)}$&${\tiny \overline{\yng(1)}}$&$\frac{4}{F-4}$&$\frac{2}{F-4}$&$0$ \\   
 $q'$&${\tiny \yng(1)}$&1&$\frac{F}{F-4}$&$\frac{2}{F-4}$&$0$ \\  
 $\bar{q}$&${\tiny  \overline{\yng(1)}}$&1&$-\frac{F}{F-4}$&$-2-\frac{2}{F-4}$&$2$  \\
$\bar{s}$ &${\tiny \overline{\yng(2)}}$&1&$-\frac{2F}{F-4}$&$-\frac{4}{F-4}$&$2$  \\
$M_{QQ}$ &1&${\tiny \yng(2)}$&2&0&0  \\
$M_{SS}$ &1&1&0&2&0  \\
$P_1$&1&${\tiny \yng(1)}$&1&2&0 \\  \hline
$Z_{SO(5)} \sim \det \bar{s}$&1&1&$-2F$&$-4$&$2F-8$  \\
$Y_{SO(7)} \sim \bar{s}^{F-5} \bar{q}^2$&1&1&$-2F$&$-8$&$2F-6$  \\
$Z_{SO(5),Q} \sim (\bar{s}q)^{F-5} \bar{q}$&1&\scriptsize 5-th anti-symm.&$5-2F$&$-4$&$2F-8$ \\ 
$P_4 \sim q^{F-4}$&1&\scriptsize 4-th anti-symm.&4&2&0  \\
$P_5 \sim q^{F-5} q'$&1&\scriptsize 5-th anti-symm.&5&2&0  \\ \hline
%$B:=Q^7$&1&7-th anti-symm. &7&0&0  \\
%$P_3:=SQ^3S$ &1&${\tiny \yng(1,1,1)}$&3&2&0  \\[5pt]
 \footnotesize $\tilde{Y}_{SU(F-6)}^{bare}$&\scriptsize $U(1)_2:$ $-4(F-6)$&1&\footnotesize $2F-\frac{4F}{F-4}$&\footnotesize $6-\frac{2F}{F-4}$&\footnotesize $-2F+14$  \\
\footnotesize $\tilde{Y}^{dressed}:= \tilde{Y}_{SU(F-6)}^{bare} \bar{s}^{F-6}$&1&1&$0$&$0$&2  \\ 
\footnotesize $Y_{SO(7)}^Q \sim \tilde{Y}^{bare}_{SU(F-6)}\bar{s}^{F-7} \bar{q}^2 \bar{s}(\bar{s}q)^{F-7} \bar{q}$&1&\scriptsize 7-th anti-symm.&$7-2F$&$-8$&$2F-6$  \\
\footnotesize $P_8 \sim  \tilde{Y}_{SU(F-6)}^{bare} (\bar{s}q)^{F-8} \tilde{W}_\alpha^2 $&1&\scriptsize 8-th anti-symm.&8&2&0  \\
\footnotesize  $B \sim  \tilde{Y}_{SU(F-6)}^{bare} (\bar{s}q)^{F-9} \bar{q} \tilde{W}_\alpha^2$&1&\scriptsize 9-th anti-symm.&9&0&0  \\ \hline
  \end{tabular}}
  \end{center}\label{Spin9magnetic}
\end{table}

%%%%%%%%%%%%%%%%%%%%%%%%%%%%%%%%%%%%%%%%%%%%%%%%%%%%%%%%%%
Let us test various consistency checks of the duality proposed above. First, we consider adding deformations to the vector meson $M_{QQ}$. Let us introduce a complex mass term for a single vector, say an $F$-th flavor as $W=m Q^FQ^F$. On the electric side, we just integrate out this vector and $F-1$ vectors remain massless. On the magnetic side, the same deformation corresponds to the higgsing with $\mathrm{rank}\braket{\bar{s}qq} =1$. The magnetic theory then reduces to an $SU(F-5)$ gauge theory with the same matter content. In this way, the duality is preserved with reduction of $F$. 
Alternatively, by introducing a vev of $\mathrm{rank} \braket{M_{QQ}}=1$, the electric theory flows to a $Spin(8)$ gauge theory with $F-1$ vectors, a spinor and a conjugate spinor. On the magnetic side, the superpotential precisely reduces to \eqref{WmagSpin8F11}. We thus flow to the $Spin(8)$ duality studied in Section \ref{Spin8F11}, which serves another check of the duality. 

Next, we consider deformations along the $M_{SS}$ direction. By introducing a non-zero vev to $M_{SS}$, the electric side flows to a 3d $\mathcal{N}=2$ $Spin(7)$ gauge theory with $F$ spinor matters. On the magnetic side, $\braket{M_{SS}} \neq 0$ corresponds to a complex mass between $q'$ and $\bar{q}$. By integrating out all the massive components, we reproduce the $Spin(7)$ duality proposed in \cite{Nii:2019wjz}. Finally, we consider adding a complex mass term as $W=m SS $ which integrates out the spinor matter in the $Spin(9)$ theory. On the magnetic side, the mass $W=m M_{SS}$ leads to the following vacuum expectation values (via an appropriate gauge transformation)
\begin{align}
q'^a&= 0,~~~~q'^{a=F-4}= \sqrt{m},~~~~\bar{q}_a =0,~~~~\bar{q}_{a=F-4}=\sqrt{m}\\
\hat{q}_i^a &:=q_i^{a},~~~~q_{i}^{a=F-4}=0\\
\hat{\bar{s}} &:= \bar{s}_{ab},~~~~\bar{s}_{singlet} := \bar{s}_{F-4,F-4},
\end{align}
where $a,b=1,\cdots,F-4$ are the color indices and $i=1,\cdots,F$ is a flavor index.
By inserting these expressions into the superpotential, we further obtain $m+ \tilde{Y}^{bare}_{SU(F-5)} \hat{\bar{s}}^{F-7} =0$ and thus find the twofold higgsing $SU(F-4) \rightarrow SU(F-5) \rightarrow S(O(F-7) \times O(2))$. The singlet $\bar{s}_{singlet} $ is decoupled from the other matter content. By dualizing the $O(2)$ vector superfield into a chiral superfield, we reproduce the $Spin(N) \leftrightarrow O(F-N+2)$ duality with $N=9$ \cite{Aharony:2013kma}. This supports the validity of our duality.

%%%%%%%%%%%%%%%%%%%%%%%%%%%%%%%%%%%%%%%%%%%%%%%%%%%%%%%%%%
%%%%%%%%%%%%%%%%%%%%%%%%%%%%%%%%%%%%%%%%%%%%%%%%%%%%%%%%%%
\section{3d $Spin(10)$ Seiberg duality}
%%%%%%%%%%%%%%%%%%%%%%%%%%%%%%%%%%%%%%%%%%%%%%%%%%%%%%%%%%
%%%%%%%%%%%%%%%%%%%%%%%%%%%%%%%%%%%%%%%%%%%%%%%%%%%%%%%%%%
As a final example of the 3d $Spin(N)$ Seiberg dualities, we consider a $Spin(10)$ gauge group with matters in vector and spinor representations. The corresponding 4d duality was studied in \cite{Pouliot:1996zh, Kawano:1996bd, Berkooz:1997bb, Kawano:2005nc} while the 3d $Spin(10)$ duality with only vector matters was proposed in \cite{Aharony:2013kma}. Here, we add a single spinor (whose representation is denoted by $\mathbf{16}$) and propose its 3d duality. The electric description is a 3d $\mathcal{N}=2$ $Spin(10)$ gauge theory with $F$ vector matters $Q$ and a spinor matter $S$. The global symmetry becomes $SU(F) \times U(1) \times U(1) \times U(1)_R$ since there is no chiral anomaly. The s-confinement phase appears when $F=6$, which was studied in \cite{Nii:2018wwj} by the author, and here we propose a dual description for $F \ge 7$. The matter content and its quantum numbers are defined in Table \ref{Spin10electric}. The two $U(1)$ symmetries just count the numbers of vectors and spinors. The Higgs branch operators are defined by
\begin{gather*}
M_{QQ}:=QQ,~~~P_1=SQS,~~~P_5 :=SQ^5S,~~~P_9:=SQ^9S,~~~B:=Q^{10},
\end{gather*}
where the mixed meson $P_5$ and $P_9$ are constructible for $F \ge 5$ and $F \ge 9$, respectively. The baryon $B$ is available only for $F \ge 10$. Notice that there is no mass operator available for a single spinor matter in $Spin(10)$. In $P_1, P_5$ and $P_9$, the color indices of two $S$'s are symmetrized. $M_{QQ}$ and $P_1$ will become elementary fields in a magnetic theory.

Next, we study the Coulomb branch operators in the $Spin(10)$ gauge theory, which was investigated in \cite{Aharony:2011ci, Aharony:2013kma, Nii:2018wwj} for theories with vector matters and for the descriptions of the s-confinement phases. The first Coulomb branch, which we denote by $Y_{SO(8)}$, leads to the spontaneous breaking of the gauge group
\begin{align}
so(10) & \rightarrow so(8) \times u(1) \\
\mathbf{10} & \rightarrow \mathbf{8_v}_0+\mathbf{1}_{2} +\mathbf{1}_{-2} \\
\mathbf{16} & \rightarrow  \mathbf{8_c}_{-1}+ \mathbf{8_s}_{1},
\end{align} 
where all the components from the spinor representation are massive along the $Y_{SO(8)}$ flat direction.  The massless component $\mathbf{8_v}_0$ from the vector matters can make the vacuum of the low-energy $SO(8)$ gauge theory stable and supersymmetric when $F \ge 7$ (no runaway potential generated) \cite{Aharony:2011ci}. Therefore, the $Y_{SO(8)}$ branch will become exactly flat and survive quantum corrections for the region of $F$ where the dual description is available.

The $Spin(10)$ gauge theory with a spinor matter can have another Coulomb branch \cite{Aharony:2013kma, Nii:2018wwj}. This second flat direction, which is denoted by $Z_{SO(6)}$, corresponds to the following gauge symmetry breaking
\begin{align}
so(10) & \rightarrow so(6) \times su(2) \times u(1) \\
\mathbf{10} & \rightarrow (\mathbf{6},\mathbf{1})_0 +(\mathbf{1},\mathbf{2})_{\pm 1}\\
\mathbf{16} & \rightarrow   (\mathbf{4},\mathbf{1})_{\pm 1}+ (\mathbf{4},\mathbf{2})_0 .
\end{align} 
The bare monopole operator $Z_{SO(6)}$ can be obtained by dualizing the unbroken $U(1)$ vector superfield into a chiral superfield. This flat direction is quantum-mechanically stable and supersymmetric because there is a massless component $(\mathbf{4},\mathbf{2})_0$ from the spinor matter and then no runaway potential is generated for both the $SO(6)$ and $SU(2)$ gauge dynamics. For $Spin(10)$ gauge theories only with vectors, this direction is unstable since the low-energy $SU(2)$ dynamics generates a runaway potential from the fundamental monopole corresponding to $U(1) \subset SU(2)$ \cite{Aharony:1997bx, deBoer:1997kr}.

Based on these two Coulomb branches, we can also define the baryon-monopoles (dressed Coulomb branch operators) by combining the Coulomb and Higgs branch chiral superfields \cite{Aharony:2013kma}
\begin{align}
Y_{SO(8)}^{Q}:=Y_{SO(8)}  (\mathbf{8_v}_0)^8   \sim Y_{SO(8)}Q^8  \\
Z_{SO(6)}^Q:=Z_{SO(6)}  \left(   (\mathbf{6},\mathbf{1})_0 \right)^6    \sim Z_{SO(6)}Q^6,
\end{align}
where the color indices of $Q^8$ and $Q^6$ are contracted by epsilon tensors of the $SO(8)$ and $SO(6)$ gauge groups, respectively. As a result, the flavor indices of $Q$'s are anti-symmetrized as well. These are regarded as a 3d version of the hybrid baryons in 4d $Spin(N)$ gauge theories \cite{Intriligator:1995id, Pouliot:1996zh, Cho:1997kr, Berkooz:1997bb, Aharony:2013kma}.

For small flavors with $F \le 6$, we find that the theory exhibits confinement phases and that the following effective superpotentials are consistent with all the global symmetries in Table \ref{Spin10electric}:
\begin{align}
W^{eff}_{F=6} &= Z_{SO(6)} \left(M_{QQ}^5  P_1^2+M_{QQ}P_5^2  \right)+  Z_{SO(6)}^Q P_1P_5 \\
W^{eff}_{F=5} &= \lambda \left[ Z_{SO(6)} (M_{QQ}^4P_1^2+P_5^2)-1 \right] \\
W^{eff}_{F \le 4} &= \left( \frac{1}{Z_{SO(6)} M_{QQ}^{F-1}P_1^2 } \right)^{\frac{1}{5-F}} ,
\end{align}
where $\lambda$ for the $F=5$ case is a Lagrange multiplier field imposing one quantum constraint on the moduli coordinates. Therefore, the origin of the moduli space is eliminated when $F=5$. For $F\le 4$, the effective superpotentials become runaway and there is no stable supersymmetric vacuum. For $F=6$, the theory exhibits s-confinement, which was studied in \cite{Nii:2018wwj}. In what follows, we are interested in the $F \ge 7$ cases where the moduli space has singularities at the origin of the moduli space and is expected to realize a non-abelian Coulomb phase.

\begin{table}[H]\caption{3d $\mathcal{N}=2$ $Spin(10)$ gauge theory with $F$ vectors and a spinor} 
\begin{center}
\scalebox{1}{
  \begin{tabular}{|c||c|c|c|c|c| } \hline
  &$Spin(10)$&$SU(F)$&$U(1)$&$U(1)$& $U(1)_R$  \\ \hline
$Q$&$\mathbf{10}$&${\tiny \yng(1)}$&1&0&$0$ \\   
 $S$&$\mathbf{16}$&1&0&1&$0$ \\  \hline 
$M_{QQ}:=Q^2$ &1&${\tiny \yng(2)}$&2&0&0  \\
%$M_{SS}:=S^2$ &1&1&0&2&0  \\
$P_1:= SQS$&1&${\tiny \yng(1)}$&1&2&0 \\
%$P_4:=SQ^4S$&1&${\tiny \yng(1,1,1,1)}$&4&2&0  \\[9pt]
$P_5:=SQ^5S$&1&\scriptsize 5-th anti-symm.&5&2&0  \\ 
$P_9:=SQ^9S$&1&\scriptsize 9-th anti-symm.&9&2&0 \\ 
$B:=Q^{10}$&1&\scriptsize 10-th anti-symm. &10&0&0  \\  \hline
%$P_3:=SQ^3S$ &1&${\tiny \yng(1,1,1)}$&3&2&0  \\[5pt]
$Y_{SO(8)}$&1&1&$-2F$&$-8$&$2F-8$  \\
$Z_{SO(6)}$&1&1&$-2F$&$-4$&$2F-10$  \\ 
$Y_{SO(8)}^{Q}:=Y_{SO(8)}Q^8$&1&\scriptsize 8-th anti-symm.&$8-2F$&$-8$&$2F-8$  \\
$Z_{SO(6)}^Q:=Z_{SO(6)}Q^6$&1&\scriptsize 6-th anti-symm.&$6-2F$&$-4$&$2F-10$  \\  \hline
%$Z_{SO(5),Q}:=Z_{SO(5)}Q^5$&1&${\tiny \yng(1,1,1,1,1)}$&$5-2F$&$-4$&$2F-8$ \\[11pt] \hline
  \end{tabular}}
  \end{center}\label{Spin10electric}
\end{table}

%%%%%%%%%%%%%%%%%%%%%%%%%%%%%%%%%%%%%%%%%%%%%%%%%%%%%%%%%%
We move on to the magnetic description which is given by a 3d $\mathcal{N}=2$ $SU(F-5)$ gauge theory with $F$ fundamental matters $q$, an anti-fundamental matter $\bar{q}$ and a symmetric-bar tensor $\bar{s}$. The theory also includes gauge singlet fields $M_{QQ}$ and $P_1$, which are straightforwardly identified with the electric counterparts in Table \ref{Spin10electric}. The field content is completely the same as the 4d one \cite{Pouliot:1996zh} and the difference only comes from $F$-terms. We introduce a tree-level superpotential to truncate the magnetic chiral ring:
\begin{align}
W_{mag} =M_{QQ}\bar{s}qq +P_1 q \bar{q} +\tilde{Y}^{dressed},
\end{align}
which completely fixes the quantum numbers of the dual elementary fields as in Table \ref{Spin10magnetic}. The definition of the dressed operator $\tilde{Y}^{dressed}$ will be stated below. Being different from the corresponding 4d duality \cite{Pouliot:1996zh}, the superpotential does not include $\det \, \bar{s}$. This fact is crucial for finding the following operator mapping:
\begin{gather*}
Z_{SO(6)} \sim \det \bar{s},~~~~Y_{SO(8)} \sim \bar{s}^{F-6} \bar{q}^2\\
P_5 \sim q^{F-5},~~~~Z_{SO(6)}^Q  \sim (\bar{s}q)^{F-6} \bar{q}.
\end{gather*}
Notice that $(\bar{s}q)^{F-5}$ is proportional to $q^{F-5} \det \bar{s}$ and cannot be an independent operator. In the 4d case, the superpotential includes $W^{4d}_{mag} \ni \det \bar{s}$ and then $\det \bar{s}$ and $\bar{s}^{F-6} \bar{q}^2$ are truncated due to an $F$-flatness condition for $\bar{s}$.

We can also match the remaining electric operators by carefully studying the magnetic Coulomb branch. When the bare Coulomb branch, denoted by $\tilde{Y}_{SU(F-7)}^{bare}$, obtains a non-zero expectation value, the gauge group is spontaneously broken as follows:
\begin{align}
SU(F-5) & \rightarrow SU(F-7) \times U(1)_1 \times U(1)_2 \\ 
{\tiny \yng(1)}  & \rightarrow  {\tiny \yng(1)}_{\, 0,-2} +\mathbf{1}_{1,F-7}+\mathbf{1}_{-1,F-7}\\
{\tiny \overline{\yng(1)}}  & \rightarrow  {\tiny \overline{\yng(1)}}_{\, 0,2} +\mathbf{1}_{-1,-(F-7)}+\mathbf{1}_{1,-(F-7)}\\
{\tiny \overline{ \yng(2)}} & \rightarrow {\tiny \overline{ \yng(2)}}_{\, 0,4} +{\tiny \overline{ \yng(1)}}_{\,-1,-F+9}  +{\tiny \overline{ \yng(1)}}_{\, 1,-F+9}  \nonumber \\
& \qquad \qquad +\mathbf{1}_{-2,-2F+14} +\mathbf{1}_{2,-2F+14}+\mathbf{1}_{0,-2F+14}   \\
\mathbf{adj.} & \rightarrow  \mathbf{adj.}_{0,0}+\mathbf{1}_{0,0}+\mathbf{1}_{0,0}+   \mathbf{1}_{2,0}+\mathbf{1}_{-2,0} \nonumber \\&\qquad \qquad  + {\tiny \yng(1)} _{\,-1,-F+5}+{\tiny \yng(1)}_{\, 1,-F+5}+{\tiny \overline{ \yng(1)}}_{\, 1,F-5}+{\tiny \overline{ \yng(1)}}_{\, -1,F-5}.
\end{align}
The Coulomb branch $\tilde{Y}_{SU(F-7)}^{bare}$ is associated with the unbroken $U(1)_1$ factor. 
Hence, the components charged under the $U(1)_1$ subgroup become massive along the $\tilde{Y}_{SU(F-7)}^{bare}$ flat direction and are integrated out from the low-energy spectrum. This results in a non-zero Chern-Simons term between $U(1)_1$ and $U(1)_2$, which turns on a non-zero $U(1)_2$ charge to $\tilde{Y}_{SU(F-7)}^{bare}$ as displayed in Table \ref{Spin10magnetic} \cite{Intriligator:2013lca}. The gauge invariant operators are constructed by considering baryon-monopole operators
\begin{align}
\tilde{Y}^{dressed}&:= \tilde{Y}_{SU(F-7)}^{bare} ( {\tiny \overline{ \yng(2)}}_{\, 0,4})^{F-7} \nonumber \\
&  \sim \tilde{Y}_{SU(F-7)}^{bare} \bar{s}^{F-7}  \\
B &:=  \tilde{Y}_{SU(F-7)}^{bare} ( {\tiny \overline{ \yng(1)}}_{\, 0,2} )^{F-10}  {\tiny \overline{ \yng(1)}}_{\, 0,2} {\tiny \overline{ \yng(1)}}_{\, 1,F-5}{\tiny \overline{ \yng(1)}}_{\, -1,F-5}  \nonumber \\
& \sim \tilde{Y}_{SU(F-7)}^{bare} (\bar{s}q)^{F-10} \bar{q}\tilde{W}_\alpha^2 \\
P_9 &:=  \tilde{Y}_{SU(F-7)}^{bare} ( {\tiny \overline{ \yng(1)}}_{\, 0,2} )^{F-9} {\tiny \overline{ \yng(1)}}_{\, 1,F-5}{\tiny \overline{ \yng(1)}}_{\, -1,F-5}  \nonumber \\
& \sim \tilde{Y}_{SU(F-7)}^{bare} (\bar{s}q)^{F-9} \tilde{W}_\alpha^2 \\
Y_{SO(8)}^{Q} &:= \tilde{Y}_{SU(F-7)}^{bare} ( {\tiny \overline{ \yng(2)}}_{\, 0,4})^{F-8} ( {\tiny \overline{ \yng(1)}}_{\, 0,2} )^2 \mathbf{1}_{0,-2F+14} ( {\tiny \overline{ \yng(1)}}_{\, 0,2} )^{F-8}  {\tiny \overline{ \yng(1)}}_{\, 0,2} \nonumber \\
& \sim  \tilde{Y}_{SU(F-7)}^{bare} (\bar{s}^{F-8} \bar{q}^2) \bar{s} (\bar{s}q)^{F-8} \bar{q},
\end{align}
where the color indices are contracted by invariant tensors of the unbroken $SU(F-7)$.
The first operator $\tilde{Y}^{dressed}$ is eliminated by the superpotential. The second and third ones are identified with $B$ and $P_5$, respectively. The fourth one is identified with the dressed Coulomb branch $Y_{SO(8)}^{Q}$.

\begin{table}[H]\caption{The $SU(F-5)$ magnetic theory dual to Table \ref{Spin10electric}} 
\begin{center}
\scalebox{0.9}{
  \begin{tabular}{|c||c|c|c|c|c| } \hline
  &$SU(F-5)$&$SU(F)$&$U(1)$&$U(1)$& $U(1)_R$  \\ \hline
$q$&${\tiny \yng(1)}$&${\tiny \overline{\yng(1)}}$&$\frac{5}{F-5}$&$\frac{2}{F-5}$&$0$ \\    
 $\bar{q}$&${\tiny  \overline{\yng(1)}}$&1&$-\frac{F}{F-5}$&$-2-\frac{2}{F-5}$&$2$  \\
$\bar{s}$ &${\tiny \overline{\yng(2)}}$&1&$-2-\frac{10}{F-5}$&$-\frac{4}{F-5}$&$2$  \\
$M_{QQ}$ &1&${\tiny \yng(2)}$&2&0&0  \\
$P_1$&1&${\tiny \yng(1)}$&1&2&0 \\  \hline
$Z_{SO(6)} \sim \det \bar{s}$&1&1&$-2F$&$-4$&$2F-10$  \\
$Y_{SO(8)} \sim \bar{s}^{F-6} \bar{q}^2$&1&1&$-2F$&$-8$&$2F-8$  \\
$P_5 \sim q^{F-5}$&1&\scriptsize 5-th anti-symm.&5&2&0  \\  
$Z_{SO(6)}^Q  \sim (\bar{s}q)^{F-6} \bar{q}$&1&\scriptsize 6-th anti-symm.&$6-2F$&$-4$&$2F-10$  \\ \hline
\small$\tilde{Y}_{SU(F-7)}^{bare}$&\scriptsize $U(1)_2:$ $-4(F-7)$&1&$2F-\frac{4F}{F-5}$&$6-\frac{2F-2}{F-5}$&$-2F+16$  \\
\small$\tilde{Y}^{dressed}:= \tilde{Y}_{SU(F-7)}^{bare} \bar{s}^{F-7}$&1&1&$0$&$0$&2  \\
\small$B \sim \tilde{Y}_{SU(F-7)}^{bare} (\bar{s}q)^{F-10} \bar{q}\tilde{W}_\alpha^2$&1&\scriptsize 10-th anti-symm. &10&0&0  \\
\small $P_9 \sim \tilde{Y}_{SU(F-7)}^{bare} (\bar{s}q)^{F-9} \tilde{W}_\alpha^2$&1&\scriptsize 9-th anti-symm.&9&2&0 \\
\scriptsize $Y_{SO(8)}^{Q} \sim \tilde{Y}_{SU(F-7)}^{bare} (\bar{s}^{F-8} \bar{q}^2) \bar{s} (\bar{s}q)^{F-8} \bar{q} $&1&\scriptsize 8-th anti-symm.&$8-2F$&$-8$&$2F-8$  \\ \hline
  \end{tabular}}
  \end{center}\label{Spin10magnetic}
\end{table}

%%%%%%%%%%%%%%%%%%%%%%%%%%%%%%%%%%%%%%%%%%%%%%%%%%%%%%%%%%
By adding complex masses to vector matters $W=m M_{QQ}$ with $\mathrm{rank} \, m=r$, the electric description flows to a $Spin(10)$ gauge theory with $F-r$ vectors and a spinor. On the magnetic side, the same mass term leads to $\mathrm{rank}  \braket{\bar{s}qq}=r$ which spontaneously breaks the gauge group into $SU(F-r-5)$ with the same matter content. Thus, the duality is preserved with reduction of $F$ to $F-r$. When $r=F-6$, the magnetic gauge group vanishes and this is consistent with the s-confinement for six vectors \cite{Nii:2018wwj}. 

By introducing a non zero vev with $\mathrm{rank}\braket{M_{QQ}}=1$, we can reproduce the $Spin(9)$ duality studied in the previous section. On the electric side, the vev for $\braket{M_{QQ}^{FF}}$ breaks the gauge group to $Spin(9)$ and a single vector is eaten via the Higgs mechanism. On the magnetic side, $\braket{M_{QQ}}$ brings the superpotential into the following form
\begin{align}
W=M_{QQ} \bar{s}qq +\bar{s}q'q' + P_1 q \bar{q} +P'_1 q' \bar{q} + \tilde{Y}^{dressed},
\end{align}
where $q'$ and $P'_1$ denote the $F$-th components of $q$ and $P_1$. By rewriting $P'_1$ as $M_{SS}$, the $Spin(9)$ duality is reproduced. 

We can also study the deformation along the moduli space labeled by $\braket{P_1} \neq 0$. On the electric side, the gauge group is higgsed into $Spin(7)$. By integrating out the massive components, the low-energy limit is described by a 3d $\mathcal{N}=2$ $Spin(7)$ gauge theory with $F-1$ spinor matters. On the magnetic side, the vev of $P_1$ corresponds to a complex mass for a single $q$ and $\bar{q}$. The resulting low-energy description becomes a 3d $\mathcal{N}=2$ $SU(F-5)$ gauge theory with $F-1$ fundamental matters and a symmetric-bar tensor with a tree-level superpotential
\begin{align}
W= M_{QQ} \bar{s}qq + \tilde{Y}^{dressed},
\end{align}
which is precisely the $Spin(7)$ spinorial duality studied in \cite{Nii:2019wjz}.

%%%%%%%%%%%%%%%%%%%%%%%%%%%%%%%%%%%%%%%%%%%%%%%%%%%%%%%%%%
%%%%%%%%%%%%%%%%%%%%%%%%%%%%%%%%%%%%%%%%%%%%%%%%%%%%%%%%%%
\section{Summary and Discussion}
%%%%%%%%%%%%%%%%%%%%%%%%%%%%%%%%%%%%%%%%%%%%%%%%%%%%%%%%%%
%%%%%%%%%%%%%%%%%%%%%%%%%%%%%%%%%%%%%%%%%%%%%%%%%%%%%%%%%%
In this paper, we proposed the 3d Seiberg dualities for the 3d $\mathcal{N}=2$ $Spin(N)$ gauge theories with vector matters and (conjugate) spinor matters. The examples range from $Spin(7)$ to $Spin(10)$ gauge groups with one or two spinor matters. The proposed dualities are almost identical to the corresponding 4d dualities except for the fact that the magnetic theory has a slightly different superpotential. In particular, the 3d magnetic superoptential includes the (dressed) monopole operators instead of $\det \, \bar{s}$. We argued that this slightly modified superpotential is very important to have the correct operator matching in 3d. As consistency checks, we discussed various deformations to the known dualities and flows between the dualities proposed here. We also calculated the superconformal indices for the $Spin(7)$ cases and observed a nice agreement under the duality. For the $Spin(8)$ case, we argued that the duality pair flows to the 3d $\mathcal{N}=4$ theories where we found that the Hilbert series of the $\mathcal{N} = 4$ pair are identical.
 
%Discussions [dressed operator]
In these dualities, some of the gauge-invariant operators whose flavor indices are anti-symmetrized are mapped to the exotic dressed monopole operators that contain the gaugino superfields $\tilde{W}_\alpha$. Since the dressed operators are defined in a semi-classical region of the Coulomb branch, these gaugino fields are massive and also obtains anomalous spins \cite{Polyakov:1988md, Dimofte:2011py} in a magnetic monopole background. (In the current case, the field $\tilde{W}_\alpha$ is equivalent to the chiral superfield with a unit r-charge in the monopole background.) From the low-energy point of view, these massive contributions must be regarded as some complicated combinations of the massless (chiral) fields. In this paper, however, we couldn't find the correct representations of these massive components in terms of the low-energy degrees of freedom although the expressions given in this paper are consistent with the viewpoint of the superconformal indices. The same problem was posed for example in \cite{Aharony:2015pla}. We will leave this problem for future directions.

%Future direction
In this paper, we couldn't find how to derive the 3d $Spin(N)$ Seiberg dualities from the 4d dualities. The main difficulty comes from the fact that we don't understand the origin of the monopole superpotential. To clarify this problem, it might be useful to study the superconformal indices in 4d (which is a partition function on $\mathbb{S}^1 \times \mathbb{S}^3$) and reduce it to 3d $\mathbb{S}^3$ partition functions \cite{Dolan:2011rp, Gadde:2011ia, Imamura:2011uw, Niarchos:2012ah}. It would be important to study more generic $Spin(N)$ Seiberg dualities by including more than one spinor matters and by studying more higher rank gauge groups. In 4d, the $Spin(10)$ gauge theory with arbitrary number of spinors was proposed in \cite{Berkooz:1997bb}.

%%%%%%%%%%%%%%%%%%%%%%%%%%%%%%%%%%%%%%%%%%%%%%%%%%%%%%%%%%
%%%%%%%%%%%%%%%%%%%%%%%%%%%%%%%%%%%%%%%%%%%%%%%%%%%%%%%%%%
\section*{Acknowledgments}
%%%%%%%%%%%%%%%%%%%%%%%%%%%%%%%%%%%%%%%%%%%%%%%%%%%%%%%%%%
%%%%%%%%%%%%%%%%%%%%%%%%%%%%%%%%%%%%%%%%%%%%%%%%%%%%%%%%%%
I would like to thank S. Sugimoto, S. Terashima, S. Yokoyama, Y. Yoshida, H. Nakajima and Y. Tachikawa for helpful discussions and comments. Keita Nii is the Yukawa Research Fellow supported by Yukawa Memorial Foundation.

%%%%%%%%%%%%%%%%%%%%%%%%%%%%%%%%%%%%%%%%%%%%%%%%%%%%%%%%%%
%%%%%%%%%%%%%%%%%%%%%%%%%%%%%%%%%%%%%%%%%%%%%%%%%%%%%%%%%%

%\bibliographystyle{unsrt}
\bibliographystyle{ieeetr}
\bibliography{3dSpindualityref}

\end{document}